\documentstyle[preprint,aps,prc,epsfig]{revtex}
\tightenlines
\newcommand{\vsig}{\mbox{\boldmath $\sigma$\unboldmath}}
\newcommand{\veps}{\mbox{\boldmath $\epsilon$\unboldmath$_\gamma$}}
\newcommand{\valf}{\mbox{\boldmath $\alpha$\unboldmath}}

\begin{document}

\title{\bf Pion photoproduction on the nucleon in the quark model}

\author{Qiang Zhao$^1$\thanks{Email address: qiang.zhao@surrey.ac.uk.}, 
J.S. Al-Khalili$^1$, Z.-P. Li$^2$, and R.L. Workman$^3$}
\address{ $^1$ Department of Physics, University of Surrey, Guildford, 
Surrey GU2 7XH,
United Kingdom\\
$^2$ Department of Physics, Peking University, 100871, Beijing, P.R. China\\
$^3$ Center for Nuclear Studies and Department of Physics, 
The George Washington University, Washington, D.C., 20052} 


\maketitle  
  
\begin{abstract}
We present a detailed quark-model study of pion photoproduction 
within the effective Lagrangian approach.
Cross sections and single-polarization observables are investigated
for the four charge channels, $\gamma p\to \pi^+ n$, 
$\gamma n\to \pi^- p$,
$\gamma p\to \pi^0 p$, and $\gamma n\to \pi^0 n$.
Leaving the $\pi N\Delta$ coupling strength to be a free parameter, 
we obtain a reasonably consistent description of these four channels
from threshold to the first resonance region. 
Within this effective Lagrangian approach,
strongly constrainted by the quark model, 
we consider the issue of double-counting which may occur if
additional {\it t}-channel contributions are included.

\end{abstract}
\vskip 1.cm



\section{Introduction}

Pion photoproduction has provided a wealth of information about 
baryon resonances. During the past three or four decades, extensive 
investigations have been carried out in both experiment and theory. 
In particular, the recent availability of high intensity electron and 
photon beams at JLab, ELSA, MAMI and ESRF has significantly 
improved the precision of pion photo- and electroproduction experiments.
A large experimental database now exists, and a significant increase
is expected once the current set of experiments has been analyzed. 

Pion photoproduction has been an important source, supplementary 
to  $\pi N$ scattering experiments, for establishing
most of the well-known baryon resonances, while providing
information on their photo-decay amplitudes. 
In the search for ``missing resonances",
other meson production channels, to which these resonances might 
have stronger couplings, are now being extensively studied 
(See e.g. Ref.~\cite{capstick-roberts} and references therein). 

Apart from a few dominant states, a considerable model-dependence
exists in resonance parameters extracted using phenomenological
approaches to the data. This has complicated the comparison with 
resonance parameters derived from quark models. 
Historically, most approaches have
adopted a factorization of the meson interaction vertices, where
the dynamical information is absorbed 
into the resonance partial-decay widths and empirical form factors. 
Consequently, parameters for the meson-nucleon-resonance couplings
and form factors have been introduced. 

Such empirical schemes
have been very important in analyses of data and the extraction of
resonance signals in pion photoproduction~\cite{walker-69}.
A number of multipole fits, taking into 
account different dynamical aspects, are now underway. 
For instance, the unitary isobar model (MAID)~\cite{drechsel-99},
containing Born terms, 5 resonances and vector meson exchanges, 
succeeds in the description of data up to 1 GeV. 
Approaches adopting constraints from fixed-$t$ dispersion 
relations are being re-visited
and applied to the Delta region~\cite{hanstein-98,aznauryan-99,crawford-nstar}.
Other approaches, using effective Lagrangians for the Delta
resonance excitation and {\it t}-channel vector-meson exchange,
can also be found in the 
literature~\cite{davidson-91,nozawa-90-1,nozawa-90-2,garcilazo-93,sato-lee}.
The SAID fits~\cite{zjli-93,arndt-96}, based on a parametrization of different
partial wave contributions, extend the analysis up to 2 GeV.
With some common features but quite different model constraints, 
these multipole fits hope to converge to a common result and obtain,
as near as possible, 
model-independent information on the resonance excitations.

There is a clear need to treat all resonances 
consistently, and to understand the relation between 
the {\it s}- and {\it u}-channel resonances and {\it t}-channel
meson exchanges. 
A recently developed quark model framework\cite{li-97}, augmented 
by an effective Lagrangian approach to reaction dynamics,
provides a good starting point.
The main feature of this model is the introduction 
of an effective chiral Lagrangian for the quark-pseudoscalar-meson
coupling in a constituent quark model.
Unlike most previous quark models, which were generally 
based on factorization of the strong interaction vertices,
the pion is treated as an elementary particle. As a result,
one can explicitly calculate the tree level diagrams for pion 
production reactions. Here, the quark model wavefunctions 
for the nucleons and baryon resonances 
provide a form factor for each interaction vertex, and
all the {\it s}- and
{\it u}-channel resonances can be consistently included.

This model has the advantage of being able to describe
a large photoproduction database, employing only 
a very limited number of parameters within a microscopic
framework. 
Applications of this model to the 
$\eta$~\cite{li-eta-95,Li:1998ni,zhao-eta,Saghai:2001yd} 
and $K$~\cite{Li:1995sia,Li:kv}  
meson photoproduction have been quite successful, and this
has motivated our study of the very extensive 
pion photoproduction database.

The quark model's well-known
underestimation of the electromagnetic (EM) transition amplitude
for the Delta resonance
makes this resonance region particularly interesting. 
As suggested in Ref.~\cite{sato-lee},
the ``bare" $\gamma N\to \Delta$ vertex could be more directly
related to the quantity given by the quark model
derivation of the Delta EM transition.
A direct examination of the Delta
excitation in $\gamma N\to \Delta\to \pi N$ might shed some light
on this question. 

This paper presents both quantitative and qualitative 
investigations of pion photoproduction.
The challenge to describe the $N^*$
resonance excitation with explicit quark and gluon 
degrees of freedom is by no means trivial,
since the correct off-shell behavior of those 
intermediate resonances is required.
Also, a clear definition of the nucleon Born terms,
associated with the gauge invariance requirement,
is essential for this effective theory. In this study,
we concentrate on the energy region 
corresponding to $E_\gamma\lesssim$ 700 MeV, where 
the role played 
by the Born terms~\footnote{We use ``Born terms" here to 
denote the amplitudes from a Born approximation, in which  
the nucleon pole terms, pion pole and contact term are included. 
In the following Sections, we use ``nucleon pole terms"
to denote the {\it s}- and {\it u}-channel nucleon exchange
amplitudes.}  and the low-lying 
resonances, in particular the $\Delta (1232)$, $S_{11}(1535)$  
and $D_{13}(1520)$, can be clarified. 
Qualitative tests have been made in order to
compare this model to a typical isobaric approach. 
Here we consider the role played by the {\it t}-channel 
vector meson exchanges in isobaric models, and the effect of neglecting
the {\it u}-channel resonance contributions.

In Section II, we outline 
the formalism aspects of our approach. In Section III, results for cross sections 
and single polarization asymmetries for the four
charge channels, $\gamma p\to \pi^+ n$, $\gamma p\to \pi^0 p$,
$\gamma n\to \pi^- p$, and $\gamma n\to \pi^0 n$ will be 
presented. The role of the {\it t}-channel vector meson exchange
will also be discussed.
Conclusions are drawn in Section IV.


\section{The model}

Before we begin a detailed analysis, a brief review of this model
is necessary.
 
\subsection{The effective Lagrangian}

For pion photoproduction, the low energy theorem (LET)
provides a crucial test near threshold. 
As shown in previous investigation by Li~\cite{li-pion},
to recover the LET, one has to
rely on the low energy QCD Lagrangian
which keeps the meson-baryon interaction invariant 
under the chiral transformation. 
Combining the low energy QCD Lagrangian with the quark model,
we introduce the quark-meson interaction 
through the effective Lagrangian~\cite{li-97}:
\begin{equation}\label{lagrangian}
{\cal L}=\overline{\psi}[\gamma_\mu(i\partial^\mu+V^\mu 
+\gamma_5 A^\mu)-m]\psi +\cdot\cdot\cdot, 
\end{equation}
where the vector and axial currents are
\begin{eqnarray}
V_\mu &=&\frac 12 (\xi^\dag\partial_\mu\xi 
+ \xi\partial_\mu\xi^\dag),\nonumber\\
A_\mu & =& i\frac 12 (\xi^\dag\partial_\mu\xi
- \xi\partial_\mu\xi^\dag), 
\end{eqnarray}
and the chiral transformation is,
\begin{equation}\label{chiral}
\xi=e^{i\phi_m/f_m}, 
\end{equation}
where $f_m$ is the decay constant
of the meson. The quark field $\psi$ in the SU(3) 
symmetry is 
\begin{equation}  
\psi =\left( \begin{array}{c}  
\psi (u)\\ \psi (d) \\ \psi (s) \end{array} \right ),   
\end{equation}  
and the meson field $\phi_m$ is a 3$\otimes$3 matrix:
\begin{equation}  
\phi_m =\left( \begin{array}{ccc}  
\frac{1}{\sqrt{2}}\pi^{0}+\frac{1}{\sqrt{6}}\eta & \pi^{+} & K^{+}\\  
\pi^{-} & -\frac{1}{\sqrt{2}}\pi^{0}+\frac{1}{\sqrt{6}}\eta & K^{0}\\  
K^{-} & \overline{K}^{0}  &-\sqrt{\frac 23}\eta
\end{array} \right ) ,
\end{equation}  
where the pseudoscalar mesons $\pi$, $\eta$ and $K$ are treated 
as Goldstone bosons. Thus, the Lagrangian in Eq.~(\ref{lagrangian})
is invariant under the chiral transformation.
Expanding the nonlinear field $\xi$ in Eq.~(\ref{chiral})
in terms of the Goldstone boson field $\phi_m$, i.e. 
$\xi=1+i\phi_m/f_m+\cdot\cdot\cdot$, 
we obtain the standard quark-meson
pseudovector coupling at tree level:
\begin{equation}\label{coupling}
H_m=\sum_{j}\frac{1}{f_m}\overline{\psi}_j\gamma_\mu^j
\gamma_5^j\psi_j\partial^\mu\phi_m \ ,
\end{equation}
where $\psi_j$ ($\overline{\psi}_j$) represents 
the $j$th quark (anti-quark) field in the nucleon.

It is still not clear whether the Goldstone bosons couple to the 
nucleon through a pseudoscalar or pseudovector coupling, or even both.
To our present knowledge, at low energies, the pseudovector coupling 
satisfies partial conservation of axial current (PCAC)
and is consistent with the LET and chiral perturbation theory 
to leading order, while the high energy study prefers 
a pseudoscalar coupling. 
As pointed out in Ref.~\cite{yaouanc}, 
the operators for the pseudoscalar and pseudovector coupling
 have the same leading order expression at quark tree level.
Therefore, Eq.~(\ref{coupling}) can be regarded as 
a reasonable starting point for investigations of pion photoproduction
in the resonance region. 

The quark-photon electromagnetic coupling is
\begin{equation}
H_e=-\sum_{j} e_j\gamma^j_\mu A^\mu({\bf k}, {\bf r}),
\end{equation}
where the photon has three-momentum ${\bf k}$, and the constituent
quark carries a charge $e_j$.

\begin{figure}
\begin{center}
\epsfig{file=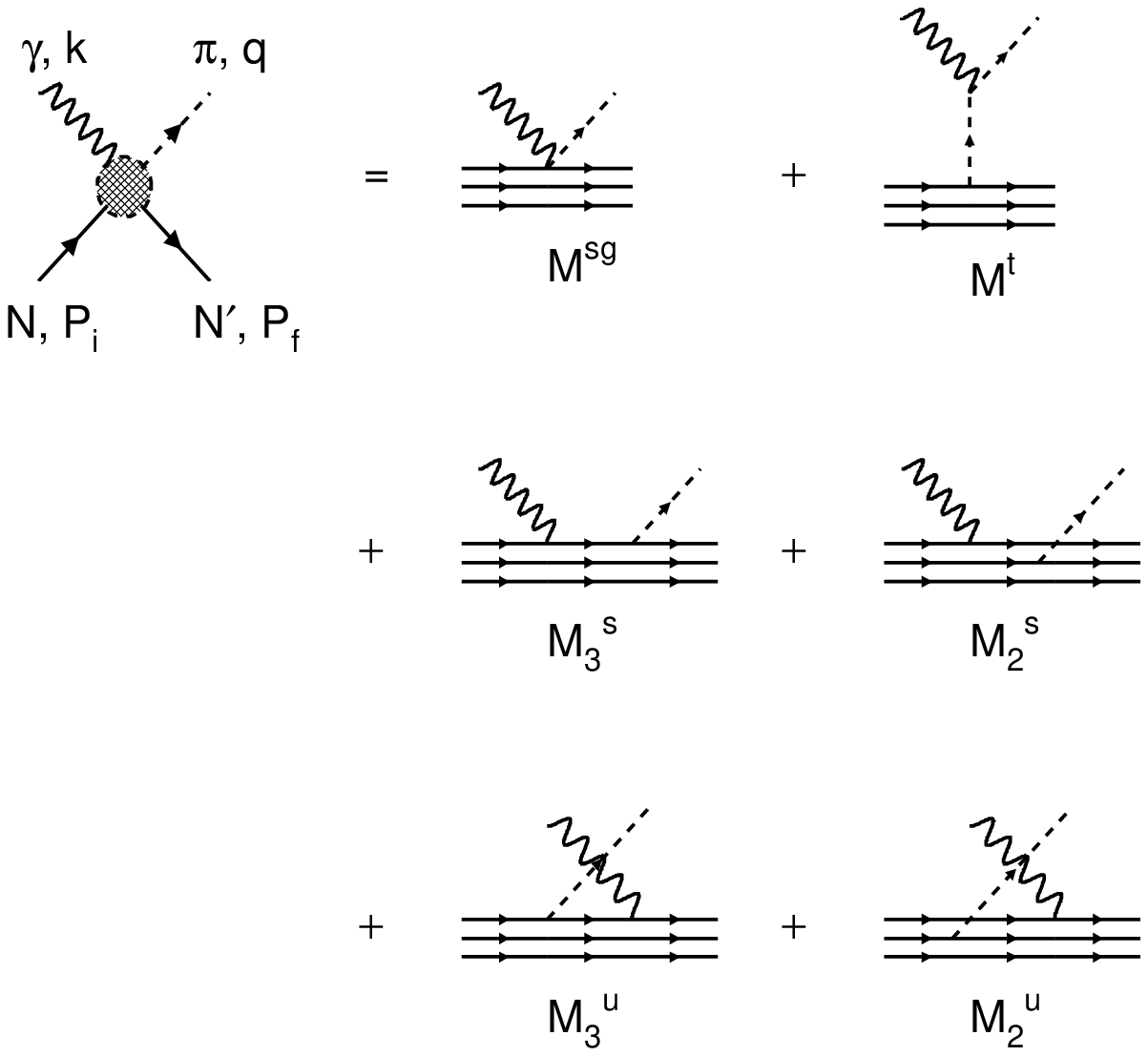,width=10cm,height=10.cm}
\end{center}
\caption{ Tree level diagrams calculated in this model.}
\protect\label{fig:(1)}
\end{figure}

The photoproduction amplitudes can be expressed in terms of 
the Mandelstam variables, 
\begin{equation}
M_{fi}=M^{sg}_{fi}+M^s_{fi}+M^u_{fi}+M^t_{fi}, 
\end{equation}
where $M^{sg}_{fi}$ is the seagull term and $M^s_{fi}$, $M^u_{fi}$
and $M^t_{fi}$ represent the {\it s}-, {\it u}-, and {\it t}-channel
processes as illustrated in Fig.~\ref{fig:(1)}. 
As shown in Ref.~\cite{li-97}, the seagull term is composed of 
two parts, 
\begin{equation}
M^{sg}_{fi}=\langle N_f |H_{m,e}|N_i \rangle 
+i\langle N_f|[g_e, H_m]|N_i \rangle, 
\end{equation}
where $|N_i \rangle$ and $| N_f \rangle$ are the initial and final state 
nucleon, respectively, and 
\begin{equation}
H_{m,e}=\sum_{j} \frac{e_m}{f_m}\phi_m({\bf q},{\bf r}_j)
\overline{\psi}_j\gamma^j_\mu\gamma^j_5
\psi_jA^\mu({\bf k}, {\bf r}_j)
\end{equation}
is the contact term from the pseudovector coupling, 
and 
\begin{equation}
g_e=\sum_j e_j{\bf r}_j\cdot\veps e^{i{\bf k}\cdot{\bf r}_j}
\end{equation}
comes from the transformation of the electromagnetic interaction
in the {\it s}- and {\it u}-channel~\cite{li-compton,li-97}. 

The {\it s}- and {\it u}-channel amplitudes have the following 
expression:
\begin{eqnarray}
&&M^s_{fi}+M^u_{fi}\nonumber\\
&=&i\omega_\gamma \sum_j\langle N_f|H_m|N_j\rangle 
\langle N_j| \frac{1}{E_i+\omega_\gamma -E_j}h_e|N_i\rangle\nonumber\\
&+&i\omega_\gamma \sum_j\langle N_f|h_e\frac{1}{E_i-\omega_m -E_j}|N_j\rangle 
\langle N_j|H_m|N_i\rangle,
\end{eqnarray}
where 
\begin{equation}
\label{em}
h_e=\sum_j e_j{\bf r}_j\cdot\veps (1-\valf_j\cdot\hat{\bf k} )
e^{i{\bf k}\cdot{\bf r}_j},
\end{equation}
and $\hat{\bf k}\equiv {\bf k}/\omega_\gamma$ is the unit 
vector in the direction of the photon momentum.

The nonrelativistic expansions of Eqs.~(\ref{em}) and ~(\ref{coupling})
become~\cite{li-97}: 
\begin{equation}
h_e=\sum_{j} \left [ e_j{\bf r}_j\cdot\veps
-\frac{e_j}{2m_j}\vsig_j\cdot(\veps\times\hat{\bf k})\right ]
e^{i{\bf k}\cdot{\bf r}_j}, 
\end{equation}
and
\begin{equation}
H^{nr}_m=\sum_j\left [\frac{\omega_m}{E_f+M_f}\vsig_j\cdot{\bf P}_f
+\frac{\omega_m}{E_i+M_i}\vsig_j\cdot{\bf P}_i
-\vsig_j\cdot{\bf q}+\frac{\omega_m}{2\mu_q}\vsig_j\cdot{\bf p}_j\right ] 
\frac{\hat{I}_j}{g_A}e^{-i{\bf q}\cdot{\bf r}_j},
\end{equation}
where $M_i \ (M_f)$, $E_i \ (E_f)$ and ${\bf P}_i \ ({\bf P}_f)$ 
are mass, energy and three-vector momentum for the initial (final) 
nucleon; ${\bf r}_j$ and ${\bf p}_j$ are the internal coordinate
and momentum for the $j$th quark in the nucleon rest system.
Note that, $g_A$, the axial vector coupling, relates the 
hadronic operator $\vsig$ to the quark operator $\vsig_j$ for the 
$j$th quark, and is defined by, 
\begin{equation}
\label{ga}
\langle N_f|\sum_j \hat{I}_j\vsig_j |N_i\rangle 
\equiv g_A\langle N_f| \vsig |N_i\rangle .
\end{equation}

The transition amplitudes of pseudoscalar meson photoproduction
can generally be expressed in terms of standard CGLN amplitudes~\cite{CGLN}, 
i.e. 
\begin{equation}
M_{fi}={\bf J}\cdot\veps, 
\end{equation}
where ${\bf J}$ is the interaction current and can be related to the 
CGLN amplitudes $f_{1,2,3,4}$:
\begin{equation}
{\bf J}=f_1 \vsig +i f_2
\frac{(\vsig\cdot{\bf q})({\bf k}\times\vsig)}{|{\bf q}||{\bf k}|} 
+ f_3\frac{\vsig\cdot{\bf k}}{|{\bf q}||{\bf k}|}{\bf q}
+f_4\frac{\vsig\cdot{\bf q}}{{\bf q}^2}{\bf q}.
\end{equation}
Alternatively, one can express the transition amplitudes
in the helicity space in terms of the ${\cal T}$ matrix:
\begin{eqnarray}
H_1&=&\langle \lambda_f=+1/2 |{\cal T}|\lambda_\gamma=+1, 
\lambda_i=-1/2\rangle\nonumber\\
H_2&=&\langle \lambda_f=+1/2 |{\cal T}|\lambda_\gamma=+1, 
\lambda_i=+1/2\rangle\nonumber\\
H_3&=&\langle \lambda_f=-1/2 |{\cal T}|\lambda_\gamma=+1, 
\lambda_i=-1/2\rangle\nonumber\\
H_4&=&\langle \lambda_f=-1/2 |{\cal T}|\lambda_\gamma=+1, 
\lambda_i=+1/2\rangle, 
\end{eqnarray}
where $\lambda_i$ and $\lambda_f$ are helicities of the initial and final 
nucleons and $\lambda_\gamma$ is the helicity of the photon. 
Amplitudes with $\lambda_\gamma=-1$ are not independent of those 
with $\lambda_\gamma=+1$ due to parity conservation.
The CGLN and helicity amplitudes 
may be related through a standard transformation~\cite{fasano92}.

\subsection{Transition amplitudes in the harmonic oscillator basis}

The seagull term in this model differs 
from the traditional definition due to the appearance 
of a transformed electromagnetic interaction coupling to the
meson at the same vertex. 
This term can be worked out explicitly in the SU(6)$\otimes$O(3)
symmetry limit:
\begin{equation}\label{sg}
M^{sg}_{fi}=-e^{-({\bf k}-{\bf q})^2/6\alpha^2}e_m
\left [1+\frac{\omega_m}{2}\left(\frac{1}{E_i+M_i}
+\frac{1}{E_f+M_f}\right)\right]
\vsig\cdot\veps,
\end{equation}
where the exponential factor
is the corresponding quark model form factor in the harmonic 
oscillator basis.

The {\it t}-channel charged pion exchange amplitude
can be derived by treating the exchanged pion as an elementary
particle:
\begin{equation}\label{t-channel}
M^t_{fi}=e^{-({\bf k}-{\bf q})^2/6\alpha^2}\frac{e_m(M_f+M_i)}{q\cdot k}
\left(\frac{\vsig\cdot{\bf q}}{E_f+M_f}-\frac{\vsig\cdot{\bf k}}{E_i+M_i}\right)
{\bf q}\cdot\veps ,
\end{equation}
where $q$ and $k$ are four-vector-momenta of the pion and photon, respectively.

As illustrated in Eqs.~(\ref{sg}) and 
(\ref{t-channel}), the leading order amplitudes 
from chiral perturbation theory are reproduced. 
The quark model modifications to these two terms
come from three-body effects, which 
add an additional term (the second term in Eq.~(\ref{sg}))
to the amplitudes. Note also the appearance of 
a form factor, which is essential to sustain 
the forward ``spike" in $\pi^+$ production.

Generalized expressions for the {\it s}- and {\it u}-channel 
amplitudes are:
\begin{equation}
M^s_{fi}=(M^s_2 + M^s_3)e^{-({\bf k}^2+{\bf q}^2)/6\alpha^2},
\end{equation}
and
\begin{equation}
M^u_{fi}=(M^u_2 + M^u_3)e^{-({\bf k}^2+{\bf q}^2)/6\alpha^2},
\end{equation}
where the $M_3$ and $M_2$ represent transitions in which
the photon and meson couple to the same quark or different 
quarks, respectively. 
The general framework was presented in Ref.~\cite{li-97}.
Here, we present the transition amplitudes
in terms of the harmonic oscillator shell $n$ as follows:
\begin{eqnarray}
\label{s3}
\frac{M^s_3}{g^s_3}&=&-\frac{1}{2 m_q}\left[ i g_v
{\bf A}_s\cdot(\veps\times{\bf k})-\vsig\cdot({\bf A}_s\times
(\veps\times {\bf k}))\right]
\frac{M_n}{n ! (P_i\cdot k -n M\omega_h )}
\left(\frac{{\bf k}\cdot{\bf q}}{3\alpha^2}\right)^{n}\nonumber\\
&+&\frac{1}{6}\left[\frac{\omega_\gamma\omega_m}{\mu_q}
(1+\frac{\omega_\gamma}{2m_q})\vsig\cdot\veps
+\frac{2\omega_\gamma}{\alpha^2}\vsig\cdot{\bf A}_s
\veps\cdot{\bf q}\right]
\frac{M_n}{(n-1) ! (P_i\cdot k -n M\omega_h)}
\left(\frac{{\bf k}\cdot{\bf q}}{3\alpha^2}\right)^{n-1}\nonumber\\
&+&\frac{\omega_\gamma\omega_m}{18\mu_q\alpha^2}
\vsig\cdot{\bf k}\veps\cdot{\bf q}
\frac{M_n}{(n-2) ! (P_i\cdot k-n M\omega_h )}
\left(\frac{{\bf k}\cdot{\bf q}}{3\alpha^2}\right)^{n-2}, 
\end{eqnarray}
and
\begin{eqnarray}
\label{s2}
\frac{M^s_2(-2)^n}{g^s_2}&=&-\frac{1}{2 m_q}\left[ i g^\prime_v
{\bf A}_s\cdot(\veps\times{\bf k})-g^\prime_a 
\vsig\cdot({\bf A}_s\times
(\veps\times {\bf k}))\right]
\frac{M_n}{n ! (P_i\cdot k-nM\omega_h)}
\left(\frac{{\bf k}\cdot{\bf q}}{3\alpha^2}\right)^{n}\nonumber\\
&+&\frac{1}{6}\left[\frac{\omega_\gamma\omega_m}{\mu_q}
(1+g^\prime_a\frac{\omega_\gamma}{2m_q})\vsig\cdot\veps
+\frac{2\omega_\gamma}{\alpha^2}\vsig\cdot{\bf A}_s
\veps\cdot{\bf q}\right]\nonumber\\
&&\times\frac{M_n}{(n-1) ! (P_i\cdot k-nM\omega_h)}
\left(\frac{{\bf k}\cdot{\bf q}}{3\alpha^2}\right)^{n-1}\nonumber\\
&+&\frac{\omega_\gamma\omega_m}{18\mu_q\alpha^2}
\vsig\cdot{\bf k}\veps\cdot{\bf q}
\frac{M_n}{(n-2) ! (P_i\cdot k-nM\omega_h)}
\left(\frac{{\bf k}\cdot{\bf q}}{3\alpha^2}\right)^{n-2}, 
\end{eqnarray}
result for the {\it s}-channel, while
\begin{eqnarray}
\label{u3}
\frac{M^u_3}{g^u_3}&=&
\frac{1}{2m_q}\left[ i g_v{\bf A}_u\cdot(\veps\times{\bf k})
+\vsig\cdot({\bf A}_u\times(\veps\times {\bf k}))\right]
\frac{M_n}{n ! (P_f\cdot k+n M\omega_h)}
\left(\frac{{\bf k}\cdot{\bf q}}{3\alpha^2}\right)^{n}\nonumber\\
&-&\frac{1}{6}\left[\frac{\omega_\gamma\omega_m}{\mu_q}
(1+\frac{\omega_\gamma}{2m_q})\vsig\cdot\veps
+\frac{2\omega_\gamma}{\alpha^2}\vsig\cdot{\bf A}_u
\veps\cdot{\bf q}\right]
\frac{M_n}{(n-1) ! (P_f\cdot k+n M\omega_h)}
\left(\frac{{\bf k}\cdot{\bf q}}{3\alpha^2}\right)^{n-1}\nonumber\\
&-&\frac{\omega_\gamma\omega_m}{18\mu_q\alpha^2}
\vsig\cdot{\bf k}\veps\cdot{\bf q}
\frac{M_n}{(n-2) ! (P_f\cdot k+n M\omega_h)}
\left(\frac{{\bf k}\cdot{\bf q}}{3\alpha^2}\right)^{n-2},
\end{eqnarray}
and
\begin{eqnarray}
\label{u2}
\frac{M^u_2(-2)^n}{g^u_2}&=&
\frac{1}{2m_q}\left[ i g^\prime_v{\bf A}_u\cdot(\veps\times{\bf k})
-g^\prime_a \vsig\cdot({\bf A}_u\times(\veps\times {\bf k}))\right]
\frac{M_n}{n ! (P_f\cdot k+n M\omega_h)}
\left(\frac{{\bf k}\cdot{\bf q}}{3\alpha^2}\right)^{n}\nonumber\\
&-&\frac{1}{6}\left[\frac{\omega_\gamma\omega_m}{\mu_q}
(1+g^\prime_a \frac{\omega_\gamma}{2m_q})\vsig\cdot\veps
+\frac{2\omega_\gamma}{\alpha^2}\vsig\cdot{\bf A}_u
\veps\cdot{\bf q}\right]\nonumber\\
&&\times\frac{M_n}{(n-1) ! (P_f\cdot k+n M\omega_h)}
\left(\frac{{\bf k}\cdot{\bf q}}{3\alpha^2}\right)^{n-1}\nonumber\\
&-&\frac{\omega_\gamma\omega_m}{18\mu_q\alpha^2}
\vsig\cdot{\bf k}\veps\cdot{\bf q}
\frac{M_n}{(n-2) ! (P_f\cdot k+n M\omega_h)}
\left(\frac{{\bf k}\cdot{\bf q}}{3\alpha^2}\right)^{n-2},
\end{eqnarray}
are corresponding terms for the {\it u}-channel. 
Vectors ${\bf A}_s$ and ${\bf A}_u$ are determined by the 
meson transitions in the {\it s}- and {\it u}-channels:
\begin{equation}
{\bf A}_s=-\left(\frac{\omega_m}{E_f+M_f}+1\right){\bf q}, 
\end{equation}
and 
\begin{equation}
{\bf A}_u=-\left(\frac{\omega_m}{E_i+M_i}+\frac{\omega_m}{E_f+M_f}
\right){\bf k}-\left(\frac{\omega_m}{E_f+M_f}+1\right){\bf q}.
\end{equation}
In Eqs.~(\ref{s3})-(\ref{u2}), $P_i$ and $P_f$ are four-vector momenta 
of the initial and final state nucleons in the total c.m. system;
$M_n$ is the mass of excited state in the $n$th shell, while 
$\omega_h$ ($=\alpha^2/m_q$) 
is the typical energy of the harmonic oscillator 
potential. The factor
$M_n/(P_i\cdot k-nM\omega_h)$ and $M_n/(P_f\cdot k+nM\omega_h)$
have clear physical meanings in the {\it s}- and {\it u}-channels
that can be related to the propagators. 

The quark level operators have been related to the hadronic
level ones through $g$-factors defined as below:
\begin{equation}
g^u_3=\langle N_f |\sum_{j} e_j\hat{I}_j\sigma^z_j|N_i\rangle /g_A,
\end{equation}
\begin{equation}
g^u_2=\langle N_f |\sum_{i\ne j} e_j\hat{I}_i\sigma^z_i|N_i\rangle /g_A,
\end{equation}
\begin{equation}
g_v=\frac{1}{g^u_3 g_A}\langle N_f|\sum_{j}e_j\hat{I}_j|N_i\rangle,
\end{equation}
\begin{equation}
g^\prime_v=\frac{1}{3g^u_2g_A}\langle N_f|\sum_{i\ne j}e_j\hat{I}_i
\vsig_i\cdot\vsig_j|N_i\rangle,
\end{equation}
and
\begin{equation}
g^\prime_a=\frac{1}{2g^u_2g_A}\langle N_f|\sum_{i\ne j} e_j\hat{I}_i
(\vsig_i\times\vsig_j)_z|N_i\rangle .
\end{equation}
Numerical values for these $g$-factors can be explicitly calculated 
in the SU(6)$\otimes$O(3) symmetry limit~\cite{li-97}.

So far, the resonance 
contributions have not been explicitly separated out. 
The intermediate states are still degenerate in the quantum number 
of the harmonic oscillator shell. 
Notice that the factor $M_n/(P_i\cdot k-nM\omega_h)$ can be written
as $2M_n/(s-M^2_n)$, where $s=W^2=(P_i+k)^2$ is the square of the 
total c.m. energy, 
we thus substitute a Breit-Wigner distribution
for the resonances, $2M_R/(s-M^2_R+iM_R\Gamma_R({\bf q}))$, 
where the resonance mass and width effects are taken into account.
Explicitly, all the contributing resonances with $n\le 2$
in the quark model symmetry limit can be included. 
In pion production, this is the place where the imaginary 
part of the transition amplitude comes out. 
The role of the imaginary part can be investigated more directly 
in the polarization observables,
e.g. polarized target asymmetry $T$ and recoil polarization
asymmetry $P$, where direct interferences between the real 
and imaginary part are highlighted.

Consequently, we must separate out the resonance excitation 
amplitudes for each $n$. 
For $n=0$, the contributing terms are
the Delta resonance excitation and the nucleon pole terms. 
One can see that only the first terms in Eqs.~(\ref{s3})-(\ref{u2})
can contribute. 
For the {\it s}-channel, we have
\begin{eqnarray}
M^s(n=0)&=&-\frac{1}{2 m_q} [ i (g^s_3g_v+g^u_2g^\prime_v)
{\bf A}_s\cdot(\veps\times{\bf k})\nonumber\\
 & &-(g^s_3+g^u_2g^\prime_a)\vsig\cdot({\bf A}_s\times
(\veps\times {\bf k})) ]
\frac{M_0}{P_i\cdot k -\delta M^2/2 }
e^{-({\bf k}^2+{\bf q}^2)/6\alpha^2},
\end{eqnarray}
where $\delta M^2/2$ 
denotes the mass square difference between the intermediate state and 
initial state nucleon.
The amplitude for the spin 1/2 nucleon pole term
is,
\begin{eqnarray}
M^s(nucleon)&=&\langle N_f|H_m|N(J=1/2)\rangle\langle N(J=1/2)|
h_e|N_i\rangle\nonumber\\
&=&-\frac{i\mu_i}{2m_q}\vsig\cdot{\bf A}_s\vsig\cdot(\veps\times{\bf k})
\frac{2M_N}{s-M^2_N}
e^{-({\bf k}^2+{\bf q}^2)/6\alpha^2},
\end{eqnarray}
where we have used $\delta M^2/2=0$ and $P_i\cdot k=(s-M^2_N)/2$ for 
the nucleon pole;
$\mu_i$ is the magnetic moment of the initial nucleon
in terms of the proton magnetic moment $e/2m_q$. 
In this way, the Delta resonance excitation amplitude
is derived,
\begin{eqnarray}
\label{delta-s}
M^s(\Delta)&=&M^s(n=0)-M^s(nucleon)\nonumber\\
&=&-\frac{1}{2 m_q} [ i (g^s_3g_v+g^u_2g^\prime_v-\mu_i)
{\bf A}_s\cdot(\veps\times{\bf k})\nonumber\\
 & &-(g^s_3+g^u_2g^\prime_a-\mu_i)\vsig\cdot({\bf A}_s\times
(\veps\times {\bf k})) ]
\frac{2M_\Delta}{s-M^2_\Delta+iM_\Delta\Gamma_\Delta}
e^{-({\bf k}^2+{\bf q}^2)/6\alpha^2},
\end{eqnarray}
where $M_0/(P_i\cdot k-\delta M^2/2)\equiv M_\Delta/(s-M^2_N-(M^2_\Delta-M^2_N))/2$
is used and
the Breit-Wigner distribution is introduced
after the separation of the spin operators.  

Similarly, the  Delta resonance and nucleon pole terms 
in the {\it u}-channel with $n=0$ can be separated:
\begin{eqnarray}
M^u(nucleon)&=&\langle N_f|h_e|N(J=1/2)\rangle\langle N(J=1/2)|
H_m|N_i\rangle\nonumber\\
&=&-\frac{i\mu_f}{2m_q}\vsig\cdot(\veps\times{\bf k})\vsig\cdot{\bf A}_u
\frac{2M_N}{u-M^2_N}
e^{-({\bf k}^2+{\bf q}^2)/6\alpha^2},
\end{eqnarray}
and 
\begin{eqnarray}
\label{delta-u}
M^u(\Delta)&=&-\frac{1}{2 m_q} [ i (g^u_3g_v+g^u_2g^\prime_v-\mu_f)
{\bf A}_u\cdot(\veps\times{\bf k})\nonumber\\
 & &+(g^u_3-g^u_2g^\prime_a-\mu_f)\vsig\cdot({\bf A}_u\times
(\veps\times {\bf k})) ]
\frac{2M_\Delta}{u-M^2_\Delta}
e^{-({\bf k}^2+{\bf q}^2)/6\alpha^2},
\end{eqnarray}
where $\mu_f$ is the magnetic moment of the final state nucleon 
in terms of the proton magnetic moment $e/2m_q$.

Several points can be learned from Eqs.~(\ref{s3})-(\ref{u2}). 
First, the nucleon pole terms and Delta resonance transition
only involve the c.m. part of their spatial wavefunctions.
Therefore, only the first terms in these equations 
contribute to the amplitudes.
For resonances with $n>0$, the internal motion of constituents
will be involved. Specifically, terms relating to $(n-1)$ 
are due to correlations between c.m. motion and internal ones, 
while $(n-2)$ terms are due to correlations 
between internal motions at two vertices.

Secondly, amplitudes for processes having the photon and 
meson coupled to different quarks are relatively suppressed.
This can be seen clearly through the factors $(-2)^n$.
In Ref.~\cite{barbour-71}, this qualitative feature 
was discussed, but not shown explicitly. Here, we show 
how the indirect diagram can be exactly calculated, 
and show that the direct diagram will become dominant 
with increasing energy and the excitation of higher states. 

Notice that in the degeneracy limit, the sum over $n$ in 
Eqs.~(\ref{s3})-(\ref{u2}) gives:
\begin{eqnarray}
M^s_{fi}+M^u_{fi}&=&\sum_{n=0}^\infty (\hat{O}_s +\hat{O}_u)
\frac{1}{n!}\left(\frac{{\bf k}\cdot{\bf q}}{3\alpha^2}\right)^n
e^{-({\bf k}^2+{\bf q}^2)/6\alpha^2}\nonumber\\
&=&(\hat{O}_s +\hat{O}_u)
e^{-({\bf k}-{\bf q})^2/6\alpha^2} \ ,
\end{eqnarray}
where $\hat{O}_s$ and $\hat{O}_u$ represent operators
independent of $n$, and recovering the 
factor $e^{-({\bf k}-{\bf q})^2/6\alpha^2}$ is essential 
for the theory to be gauge invariant. 
Although like many other phenomenological approaches
this model does not have unitarity,
such a form factor prevents certain terms from violating unitarity.
One can see that
at high energies the degeneracy limit can be 
recognized by the dominance of the direct diagram.
 How to restore the unitarity in a general 
framework should be a next step of this investigation.

\begin{figure}
\begin{center}
\epsfig{file=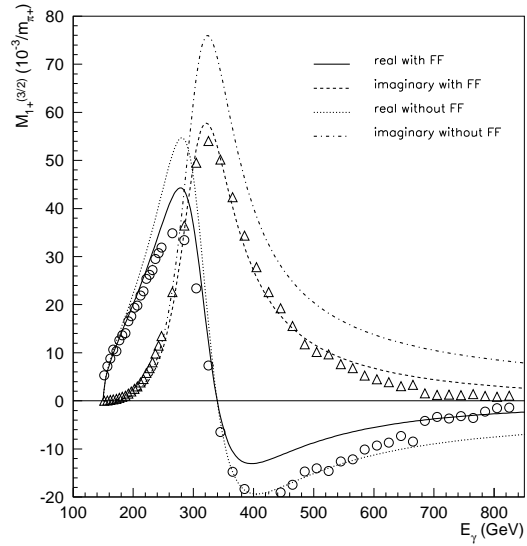,width=8cm,height=8.cm}
\end{center}
\caption{ Real and imaginary parts of the multipole $M_{1+}^{3/2}$
derived from the Delta amplitude in $\pi^0$ channel. Model results
{\it with} and {\it without} the quark model form factor (FF)
are compared.
Data are from SAID analysis~\protect\cite{pion-data}.}
\protect\label{fig:(2)}
\end{figure}

From Eq.~(\ref{delta-s}), the analytical expression
for the Delta multipole can be derived, 
\begin{equation}
M_{1+}^{3/2}=-g_{\pi NN}g_R\frac{1}{2m_q}
\left[\frac{\omega_m}{E_f+M_f}+1\right]
\frac{2M_\Delta}{s-M^2_\Delta+iM_\Delta\Gamma_\Delta}
e^{-({\bf k}^2+{\bf q}^2)/6\alpha^2},
\end{equation}
where $g_R\equiv g^s_3g_v+g^u_2g^\prime_v-\mu_i$, and 
$g_{\pi NN}$ has been taken into account. 
The real and imaginary parts of the Delta multipole
$M_{1+}^{3/2}$ are calculated and shown in Fig.~\ref{fig:(2)}. 
We shall discuss the quark model form factor effects 
in the following Section. Therefore, it would be useful to present 
the calculation of $M_{1+}^{3/2}$ {\it without} 
the exponent, which comes from the spatial integral and serves
as a form factor for the multipole. 
As illustrated by the dotted and dot-dashed curve in Fig.~\ref{fig:(2)}, 
apparent deviations from the experimental data occur with the increasing 
energies. In another word, the quark model  
turns to be indispensible to account for the correct energy dependence.

Multipole $E_{1+}^{3/2}$ vanishes in this approach 
for the {\it s}-channel Delta resonance. Experimentally, 
$E_{1+}^{3/2}$ is found to be much smaller 
than $M_{1+}^{3/2}$~\cite{arndt-96,hanstein-98}.

\section{Results and analysis}

In this Section, we present our study of 
reactions, $\gamma p\to \pi^+ n$, $\gamma p\to \pi^0 p$, 
$\gamma n\to \pi^- p$, and $\gamma n\to \pi^0 n$
with the same set of coherent parameters. 
The Goldberg-Treiman relation, 
\begin{equation}
\label{Gold-Trei}
g_{\pi NN}=\frac{g_A M_N}{f_\pi},
\end{equation}
relates the axial vector coupling $g_A$ to the well-known 
$g_{\pi NN}$ coupling, 
where $f_\pi$ is the pion decay constant. 
Note that $g_A$ in this model is an overall constant,
and can be calculated in the quark model. 
However, the quark model predicts a rather large value: 
$|g_A|=5/3$ for charged pions and $5/3\sqrt{2}$ for neutral pions.
Consequently, 
$g_{\pi NN}$ given by Eq.~(\ref{Gold-Trei}) is not a good input for our purpose.
On the other hand, $g_{\pi NN}$ is a well-determined number,
and we shall therefore ``fix" the ``parameter" $g_{\pi NN}=13.48$  
in our calculations. This is a strong 
constraint on the theory, as the 
Born terms (including seagull term, 
{\it t}-channel pion exchange, and nucleon pole terms) are completely 
fixed.  We shall follow the quark model predictions
for relative signs and strengths in order to
study the four charge channels coherently.

\subsection{The charged pion production reaction}

A distinguishing feature of $\pi^+$ photoproduction
is the forward ``spike" and the dip near $-t=m^2_\pi$ seen
in its differential cross sections.
Multipole analyses based on a unitary isobaric model
and studies utilizing fixed-t dispersion relations suggest that 
this feature is related to an interference between 
the Born terms
and the Delta resonance
``excitation" (in the naive quark model, the Delta
resonance is the ground state for isospin 3/2 baryons).
The $M^{3/2}_{1+}$ multipole for the Delta resonance 
dominates the
cross sections and single polarization asymmetries
over a wide energy range. 
Reproducing this combination of features is non-trivial 
in a microscopic model.

\begin{figure}
\begin{center}
\epsfig{file=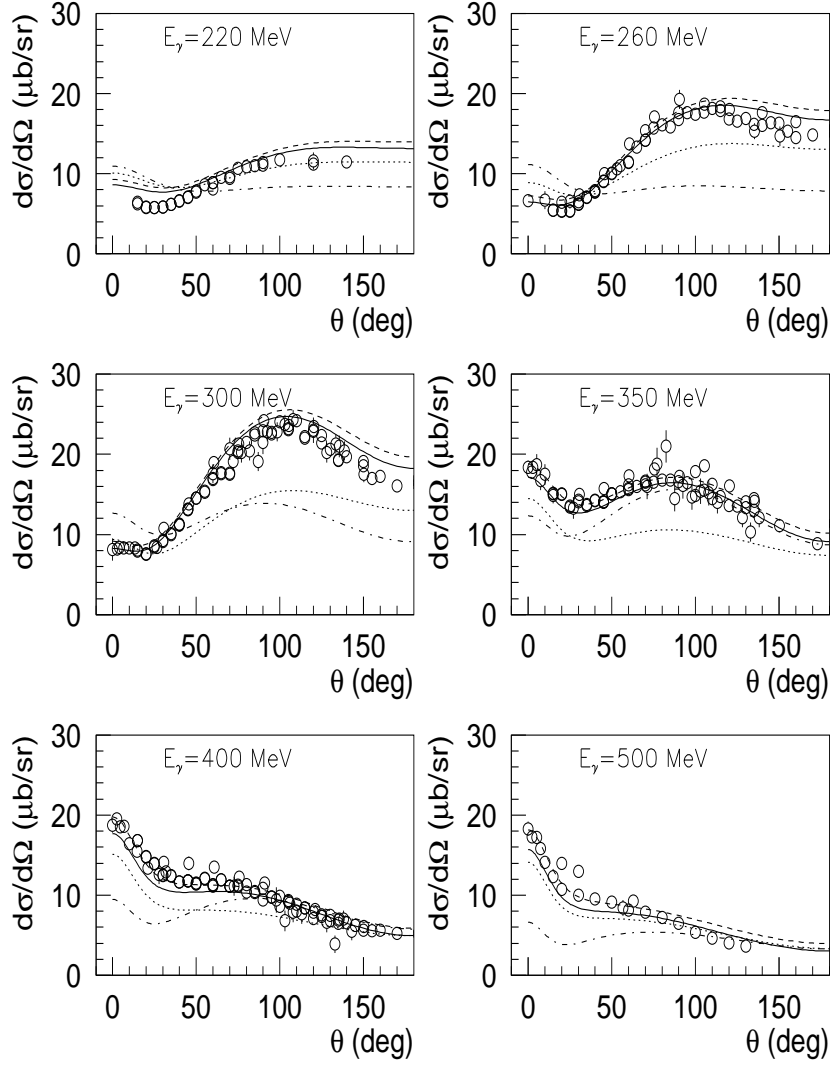,width=13.cm,height=16.cm}
\end{center}
\caption{ Differential cross sections for $\gamma p\to \pi^+ n$.
The solid curves denote full calculations 
with $C_{\Delta}=1.7474$;
Dotted curves, results in the SU(6)$\otimes$O(3) symmetry limit;
Dashed curves, results without $n>1$ {\it u}-channel resonance contributions;
Dot-dashed curves, calculations with resonance real parts eliminated.
Data are from Ref.~\protect\cite{bu(94),fi(70),fi(72),kn(63),bt(68),fu(77),br(00),al(83)}. }
\protect\label{fig:(3)}
\end{figure}

In Fig.~\ref{fig:(3)}, we show the calculations 
for reaction $\gamma p\to \pi^+ n$ in the 
SU(6)$\otimes$O(3) symmetry limit. 
The dotted curves are calculated in this limit,
having only an overall quark-meson coupling parameter,
which is in fact not ``free" for the Born terms
due to the 
Goldberg-Treiman relation. 
Near threshold, the cross section 
is reproduced reasonably, 
which is consistent with the leading order calculation
of chiral perturbation theory.
The dip at $-t= m_\pi^2$ clearly 
originates from the interference between the 
{\it t}-channel pion exchange and the seagull term.
At $E_\gamma \le 220$ MeV, 
the Delta resonance has only small interference.

\begin{figure}
\begin{center}
\epsfig{file=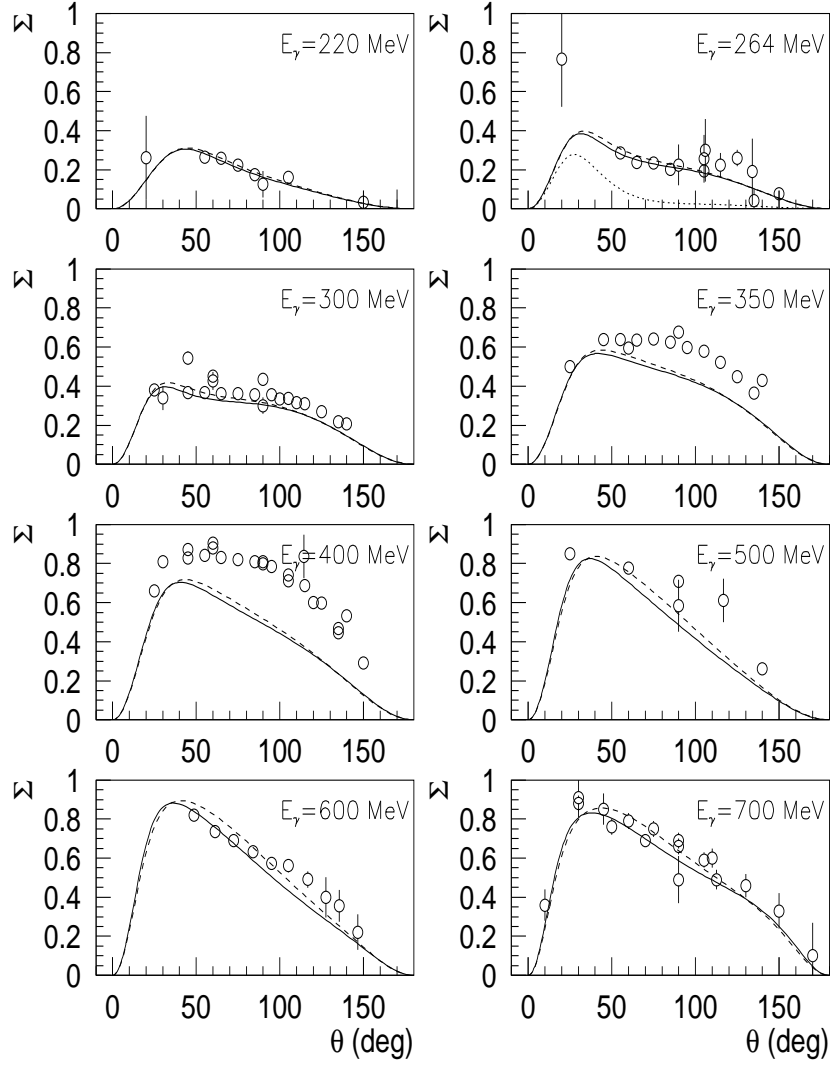,width=13.cm,height=16.cm}
\end{center}
\caption{ Polarized beam asymmetry $\Sigma$ for $\gamma p\to \pi^+ n$.
The solid curves denote full calculations, while the dashed present
results without $n>1$ {\it u}-channel resonance contributions. 
The dotted curve at $E_\gamma=260$ MeV denotes the effects
by eliminating the nucleon pole terms.
Data are from Ref.~\protect\cite{bl(01),za(75),ge(81),gn(76),lu(66),aj(00),ke(74),bs(79),as(72)}. }
\protect\label{fig:(4)}
\end{figure}

Interference from the nucleon pole terms 
becomes important from $E_\gamma=220$ to 260 MeV. 
At the lower energy limit, the {\it t}-channel pion exchange
and the seagull term dominate over other processes, while 
for higher energies, the Delta resonance dominates. 
The influence of the nucleon pole terms can be seen 
clearly in the polarized beam asymmetries.
As shown by the dotted curve in Fig.~\ref{fig:(4)} at $E_\gamma=260$ MeV,
for which the nucleon pole terms are eliminated, 
the interference from the nucleon pole terms generally
produces large asymmetries at intermediate angles.

With increasing energy, we find that the cross sections
are underestimated over the Delta resonance region in the
SU(6)$\otimes$O(3) symmetry limit.  
This suggests a failure of the symmetric quark model 
for the Delta resonance. 
However, notice that the ``dip" feature is still sustained 
over this region, and we assume that the Delta excitation
has a ``good" form factor from the quark model. Thus, 
we empirically treat the $\pi N\Delta$ coupling strength as 
a free parameter, which will be fixed by the experimental data.
The solid curves shown in Fig.~\ref{fig:(3)} 
denote the calculations with $C_\Delta=1.7474$, where
$C_\Delta=1$ is the strength in the quark model symmetry limit. 
The enhancement of the Delta contribution significantly
improves the description of the experimental data. 
Compared with the dotted curves, 
the cross sections at the extremely-forward
and intermediate angles are both enhanced.

The differential cross section for the $\pi^+$ production 
changes rapidly from threshold to the Delta resonance region. 
After that, it remains stable up to $E_\gamma\approx$ 700 MeV,
where resonances of the second resonance region start to interfere.
The challenge for a microscopic approach is not only 
to reproduce the dramatic changes at low energies, but also 
to sustain the forward peaking to high energies.
It is quite natural for us to achieve the first goal
in this model. That is, the enhanced Delta resonance 
succeeds in reproducing the drastic change of the cross sections
at the first resonance region. For the second goal, we find that 
with only one parameter, the strong forward peaking 
can only be sustained up to $E_\gamma\approx $ 500 MeV.
This result has non-trivial implications. 
It suggests that the quark model within an effective Lagrangian
provides correct signs and even reasonable form factors 
for the Delta excitation and nucleon pole terms. 
As illustrated in Section II, the Delta excitation
and the nucleon pole terms have simple structures coming from the harmonic 
oscillator shell $n=0$.  Clearly, a self-consistent
treatment\cite{workman-nstar} of these ingredients is 
essential to any viable model. 
We suggest that the tree level calculation, 
based on the quark model,
may have included the main ingredients, 
(e.g. relative signs and form factors),
even though its description
of the nucleon pole and Delta resonance is very simple.

An interesting question arising in this work 
is the role played by the {\it u}-channel resonance contributions. 
Generally, this part has been neglected in isobaric models, 
nor is it included in
traditional quark model calculations,
due to empirical considerations~\cite{barbour-71}.
In the present calculation, we find that
the {\it u}-channel process 
tends to decrease the forward peaking. 
In Fig.~\ref{fig:(3)}, the dashed curves
denote calculations with the {\it u}-channel of $n>1$ 
neglected, which enhances the forward peak above the Delta resonance region. 
Since the full calculation underestimates the forward peaking
slightly, the neglect of the {\it u}-channel of $n>1$ 
seems to follow the data more closely. 
This feature seems consistent with 
findings of Ref.~\cite{drechsel-99}. 
There, the {\it u}-channel 
resonance contributions were neglected, and an overall 
strong forward peaking was observed.

Polarization observables are sensitive 
to resonance contributions, providing a possible 
way to clarify the role played by the Delta resonance.
In Fig.~\ref{fig:(4)}, the polarized beam asymmetry $\Sigma$
is calculated for 8 energy bins. 
The results are generally in agreement
with the data at $E_\gamma\le $ 300 MeV.
However, some discrepancies are found at $E_\gamma=350$ and 400 MeV,
which are sensitive to the {\it u}-channel nucleon pole 
rather than the {\it u}-channel resonances ($n>1$).
As shown by the dashed curves, neglecting 
the $n>1$ {\it u}-channel resonance does not change the solid curves
significantly.  At $E_\gamma\approx$ 700 MeV, 
the $S_{11}(1535)$ becomes a strongly interfering source. 
The enhancement of this asymmetry at $\theta= 130$-$140^\circ$
is evidence for the existence of the $S_{11}(1535)$ resonance.

The presence of the $S_{11}(1535)$ as a state of representation 
$[{\bf 70}, \ ^{\bf 2} {\bf 8}]$ in the quark model
accounts for $\Sigma$ naturally up to 750 MeV. 
Compared with the precise measurement of GRAAL~\cite{aj(00)}, 
we cannot produce the structure observed at backward angles above
800 MeV. As suggested by the isobaric approach~\cite{drechsel-99}, 
a small $S_{11}(1650)$ contribution can reproduce the data reasonably. 
In our model calculation, 
the $S_{11}(1650)$ is absent in the proton reaction 
due to the Moorhouse selection rule~\cite{moorhouse}.
The breaking of the symmetric quark model will introduce 
mixing between states of different representations, e.g.
the $S_{11}(1535)$ and $S_{11}(1650)$. A more realistic 
model taking into account such a mixing mechanism is clearly required
above the second resonance region. 
The $S_{11}(1650)$ has large branching fraction
to $\pi N$ states~\cite{PDG2000}.

\begin{figure}
\begin{center}
\epsfig{file=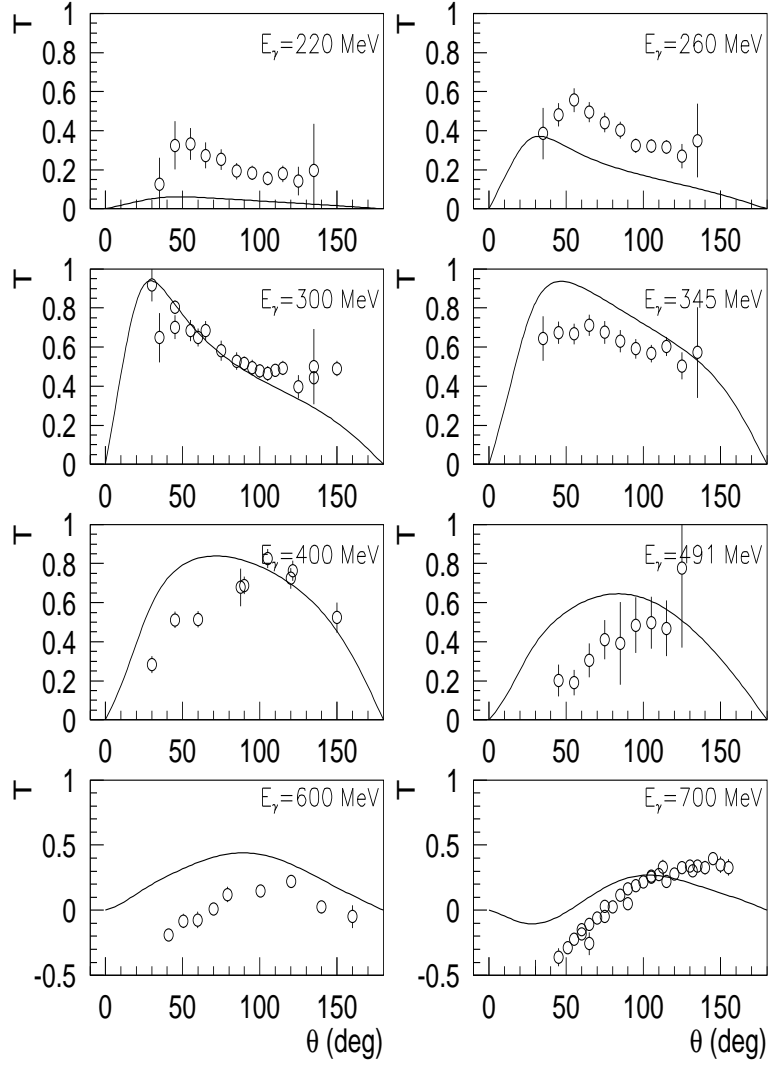,width=12.cm,height=16.cm}
\end{center}
\caption{ Polarized target asymmetry $T$ for $\gamma p\to \pi^+ n$.
Data are from Ref.~\protect\cite{du(96),ge(81),aa(72),fk(77),al(75)1}. }
\protect\label{fig:(5)}
\end{figure}

Calculations for the polarized target asymmetries 
are presented in Fig.~\ref{fig:(5)}, and compared with 
existing experimental measurements.
Given that only one parameter has been introduced, 
the results should be regarded as consistent with the data
from threshold to $E_\gamma\approx$ 500 MeV. 
At $E_\gamma=220$ MeV, our results underestimate the data, 
however this feature is 
consistent with the SAID fit~\cite{arndt-96}.

\begin{figure}
\begin{center}
\epsfig{file=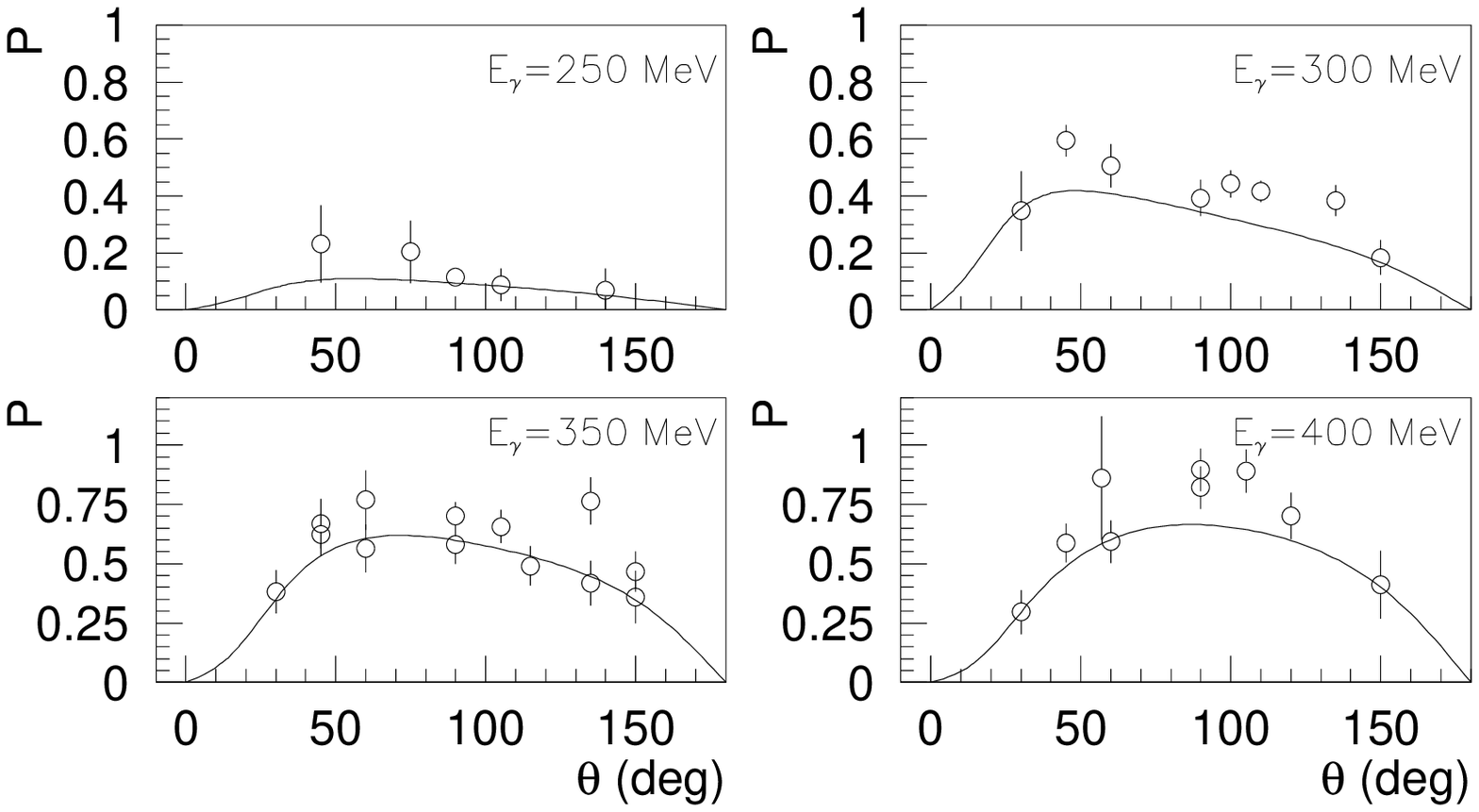,width=13.cm,height=16.cm}
\end{center}
\caption{ Recoil polarization asymmetry $P$ for $\gamma p\to \pi^+ n$.
Data are from Ref.~\protect\cite{ge(81),al(86)}.}
\protect\label{fig:(6)}
\end{figure}

Calculations for the recoil polarization asymmetries 
are presented in Fig.~\ref{fig:(6)}, which are consistent
with the data in the first resonance region.

\begin{figure}
\begin{center}
\epsfig{file=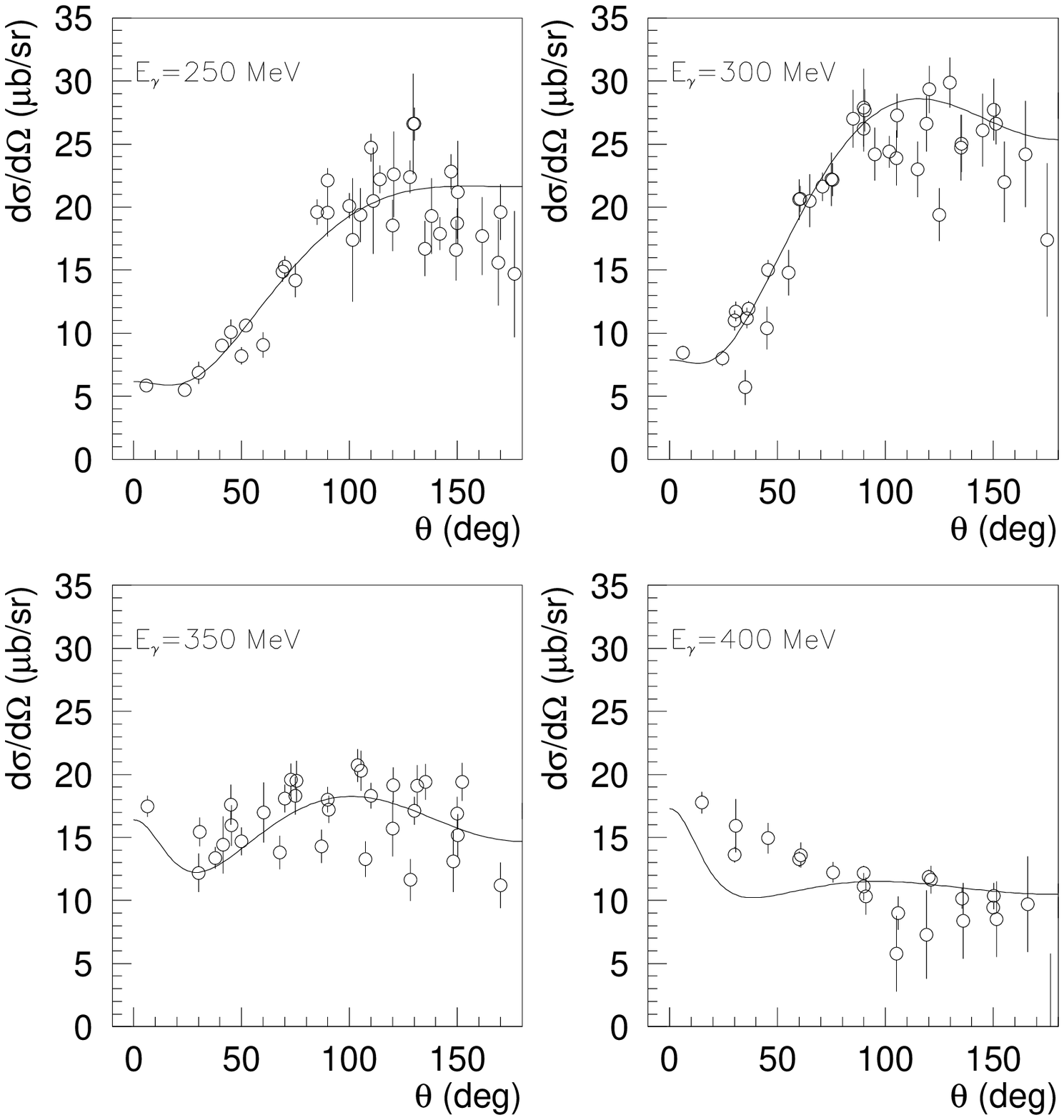,width=12.cm,height=12.cm}
\end{center}
\caption{ Differential cross sections for $\gamma n\to \pi^- p$.
Data are from Ref.~\protect\cite{fu(77),ba(88),ro(73),tr(79),ag(78),be(74),be(73),ho(74)}.}
\protect\label{fig:(7)}
\end{figure}

The reaction $\gamma n\to \pi^- p$ is calculated using 
the same set of parameters determined in the $\pi^+$ production.
We present the results for the differential cross sections
in Fig.~\ref{fig:(7)}. 
Although large uncertainties exist within the data,
our calculation is in good agreement with  experiment
up to $E_\gamma\approx $ 400 MeV. Interestingly, 
these results, which can be regarded as predictions
of this approach, are very close to the 
analyses of Ref.~\cite{hanstein-98}. It is worth noting
that similar structures as found 
in $\gamma p\to\pi^+ n$ (the ``dip" and ``spike")
are also present here, and due to the same mechanism.

\subsection{The neutral pion production reaction}

In comparison with the charged pion production, 
the neutral-pion channels are relatively simple
in this model. The contact term and the {\it t}-channel pion 
exchange are eliminated in the effective interaction
since these amplitudes are proportional to the charge
of the produced meson. The nucleon pole terms and the Delta
excitation therefore dominate over other processes 
near threshold.

\begin{figure}
\begin{center}
\epsfig{file=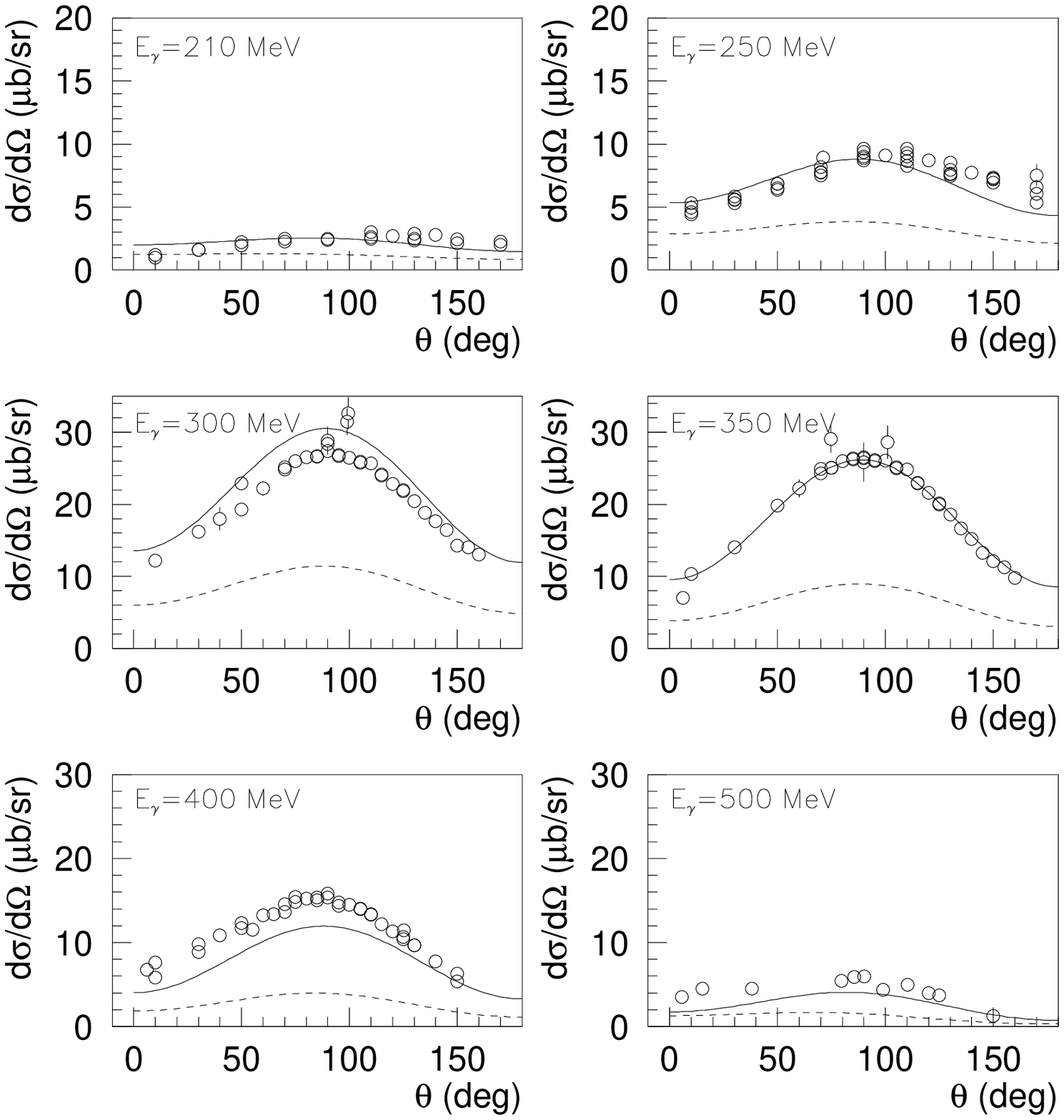,width=13.cm,height=14.cm}
\end{center}
\caption{ Differential cross sections for $\gamma p\to \pi^0 p$.
Data are from Ref.~\protect\cite{fu(96),gz(74)1,be(97),ak(78),he(73),do(77),ag(75),yo(80),hy(73)}.}
\protect\label{fig:(8)}
\end{figure}

In Fig.~\ref{fig:(8)}, the differential cross sections
for $\gamma p\to \pi^0 p$ are presented at several energies. 
In the SU(6)$\otimes$O(3) symmetry limit, as shown by the dashed curves,
the cross sections are underestimated by the Delta excitation. 
Similar to the feature arising from the charged pion production
channels, we need to enhance the Delta excitation strength
to reproduce the data. 

\begin{figure}
\begin{center}
\epsfig{file=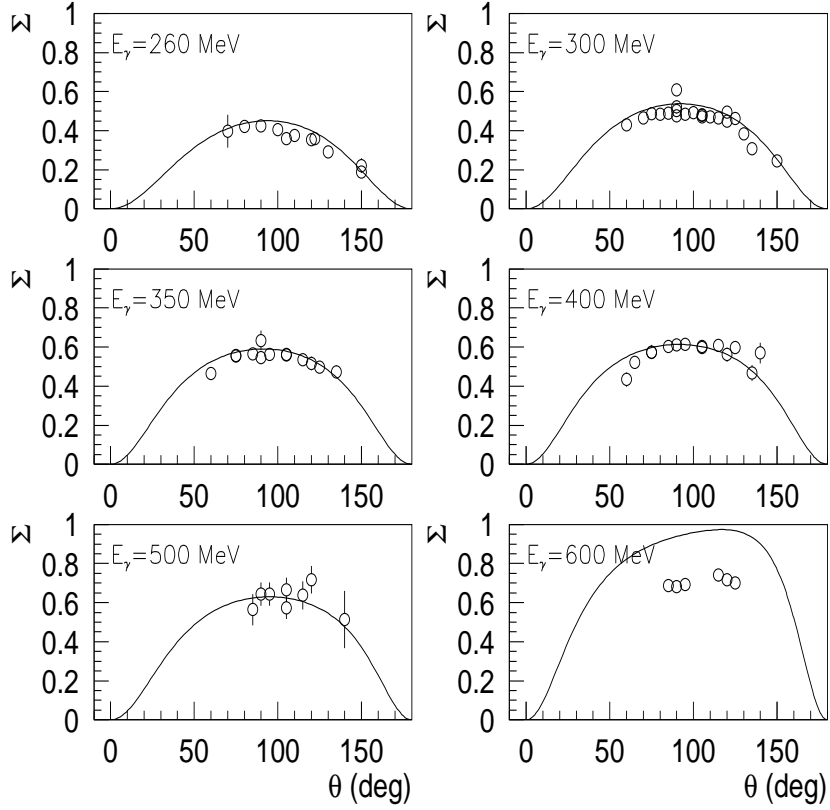,width=13.cm,height=16.cm}
\end{center}
\caption{ Polarized beam asymmetry $\Sigma$ for $\gamma p\to \pi^0 p$.
Data are from Ref.~\protect\cite{bl(01),bl(92),be(97),bl(83),bj(69),gb(78),ad(01),gb(77)}.}
\protect\label{fig:(9)}
\end{figure}

\begin{figure}
\begin{center}
\epsfig{file=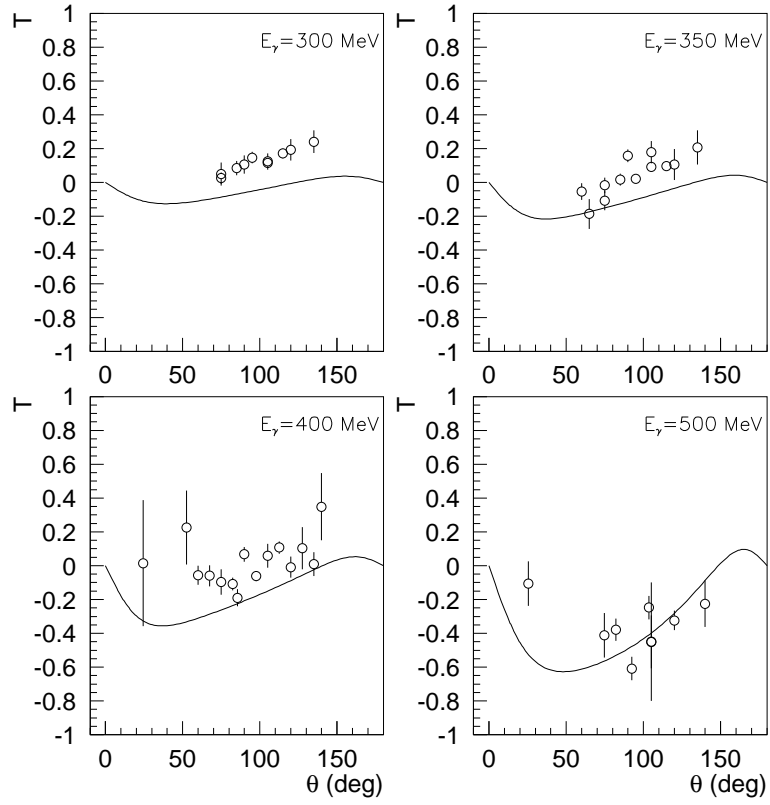,width=12.cm,height=12.cm}
\end{center}
\caption{ Polarized target asymmetry $T$ for $\gamma p\to \pi^0 p$.
Data are from Ref.~\protect\cite{bl(83),bo(98),fk(78),fe(76),gb(77)}.}
\protect\label{fig:(10)}
\end{figure}

\begin{figure}
\begin{center}
\epsfig{file=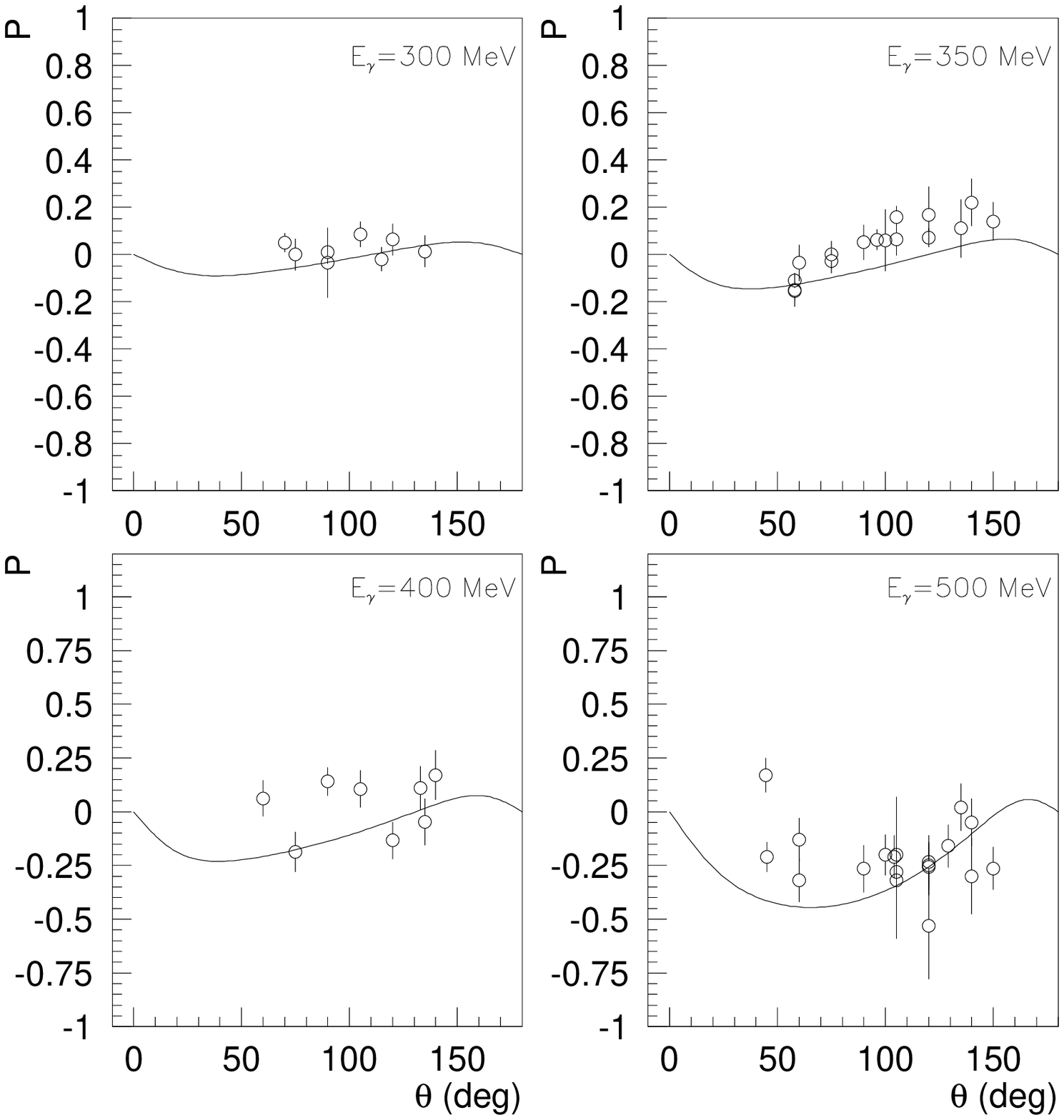,width=12.cm,height=12.cm}
\end{center}
\caption{ Recoil polarization asymmetry $P$ for $\gamma p\to \pi^0 p$.
Data are from Ref.~\protect\cite{al(68),bl(83),gb(78),gb(77)}.}
\protect\label{fig:(11)}
\end{figure}

The calculations of the single polarization asymmetries are 
presented in Fig.~\ref{fig:(9)}-~\ref{fig:(11)}
and compared with data.

\begin{figure}
\begin{center}
\epsfig{file=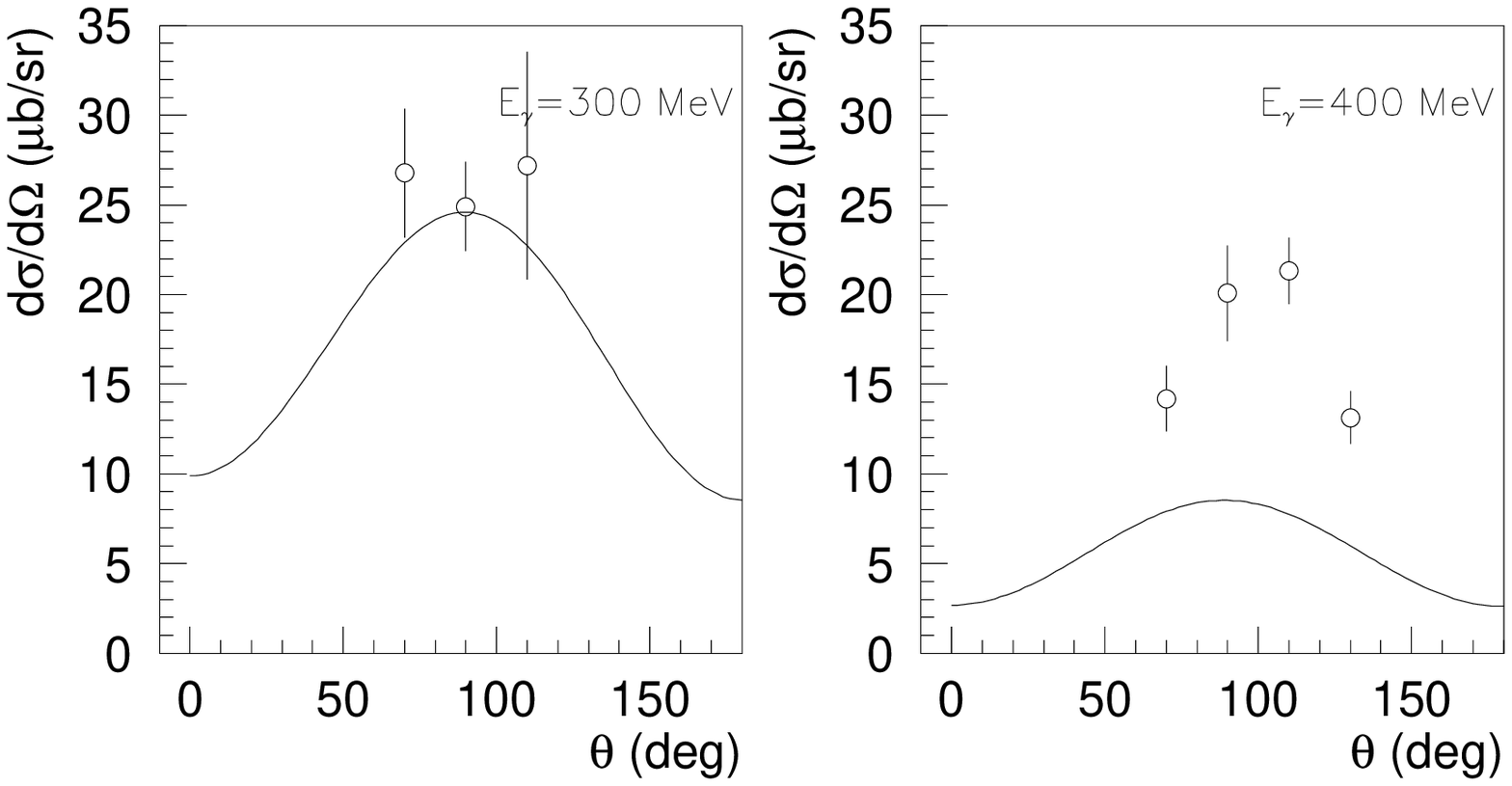,width=12.cm,height=12.cm}
\end{center}
\caption{ Differential cross sections for $\gamma n\to \pi^0 n$.
Data are from Ref.~\protect\cite{an(76)}.}
\protect\label{fig:(12)}
\end{figure}

In Fig.~\ref{fig:(12)}, the calculated cross sections
for $\gamma n\to \pi^0 n$ are presented. So far, there are only sparse 
data available for this channel.

The above results 
for the cross section and single polarization
asymmetries suggest
an overall quantitative agreement with the data 
from threshold to the first resonance region,
while qualitatively, the data up to the second resonance
region can be explained. The lack of quantitative agreement
above the Delta resonance region was expected, given that only a minimum
number of free parameters are used here. However, by using only a 
minimal model, it has been easier to identify key ingredients responsible
for those trends we {\it do} reproduce.

\subsection{Quark model form factor and the helicity basis}

As mentioned previously, in $\pi^+$ photoproduction,
the most prominent features seen 
in the cross section are forward peaking
and the dip at $-t= m^2_\pi$, which is attributed to
the Born terms. Our results also reproduce this feature. 
Some new ingredients appearing 
in this approach concern the roles played by the Born terms
and the Delta resonance, and the influence of their
associated form factors.

As found in previous studies, the Born terms 
deviate significantly from the experimental
data at intermediate and backward angles as photon energies 
increase to the GeV level.  The cross section due to 
Born terms alone is much larger than the data suggest.
One possible explanation is that
the Born terms are cancelled by resonance contributions away from
the forward peak.  As discussed by Barbour, Malone and Moorhouse
in a fixed-$t$ dispersion relation~\cite{barbour-71},
the real parts of the resonance amplitudes tend to 
cancel the Born terms at $-t>m^2_\pi$, while the region
$-t<m^2_\pi$ is slightly enhanced 
by low-lying resonance contributions. 

\begin{figure}
\begin{center}
\epsfig{file=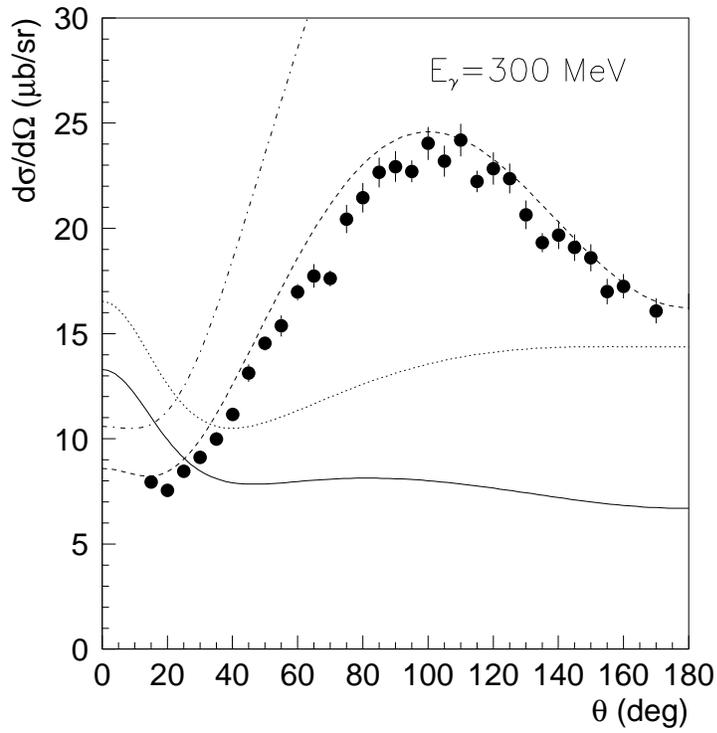,width=12.cm,height=12.cm}
\end{center}
\caption{ Cross sections for $\gamma p\to \pi^+ n$ at 300 MeV.
Plotted are the Born terms with (solid curve) and
without (dotted) form factor, and for the Born terms plus the Delta
transition with (dashed) and without (dot-dashed) form factors, respectively.}
\protect\label{fig:(13)}
\end{figure}

In Fig.~\ref{fig:(13)}, we illustrate the results
for the Born terms and Born terms plus Delta excitation,
with and without the quark model form factors, respectively.
Clearly, form factors are vital in the quark-model description,
though no free parameters have been introduced.
Comparing the full result to one in which the form factors are
switched off, we see potential problems for those who compare
quark-model results directly to fits (such as SAID and MAID)
which do not introduce form factors. 
An interesting extension of this work would be the consideration of forward
cross sections at higher energies, where the influence of 
form factors is nebulous~\cite{davidson-workman-01,workman-nstar}. 

\begin{figure}
\begin{center}
\epsfig{file=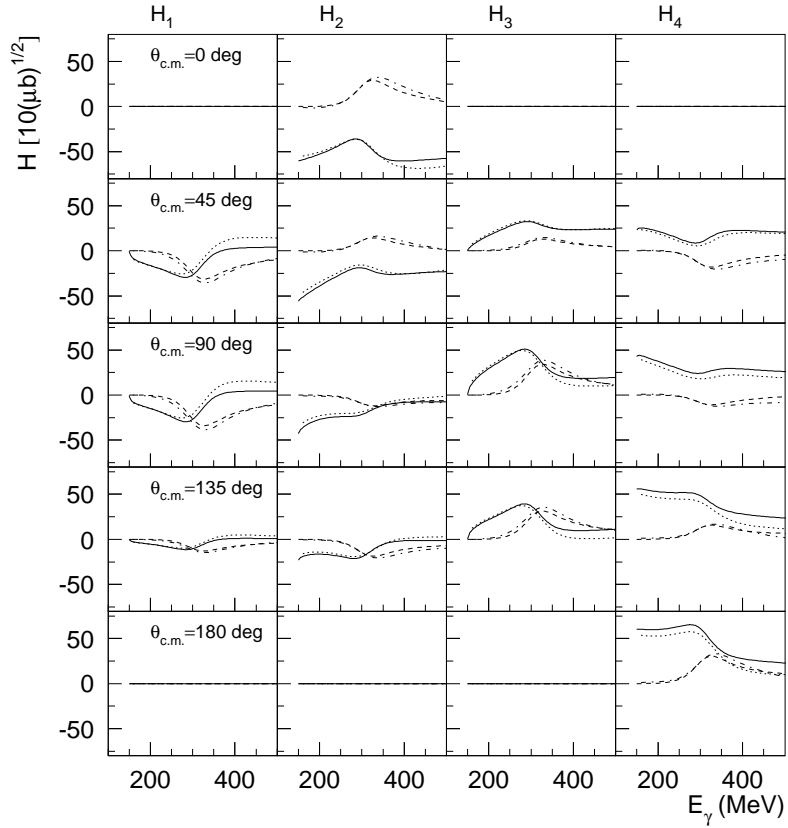,width=12.cm,height=12.cm}
\end{center}
\caption{ Energy evolution of the Born terms plus Delta helicity amplitudes 
compared with the SAID analyses. 
Columns from left to right 
are amplitudes $H_{1,2,3,4}$. The solid and dashed curves denote
the real and imaginary part calculated by this model,
while the dotted and dot-dashed denote those by SAID analyses.
 }
\protect\label{fig:(14)}
\end{figure}

To end this Section, we present a comparison of 
energy evolution of the Born terms plus Delta helicity amplitudes
calculated by this model with a SAID analysis\cite{born-delta}. 
The four independent
helicity amplitudes are calculated following the convention of 
Ref.~\cite{walker-69},
\begin{equation}
\frac{d\sigma}{d\Omega_{cm}}=\frac 12\frac{|{\bf q}|}{|{\bf k}|}
\sum_{i=1}^{4}|H_i(\theta_{cm})|^2 \ ,
\end{equation}
where $\theta_{cm}$ is the angle between the incoming photon momentum ${\bf k}$
and outgoing meson momentum ${\bf q}$ in the c.m. system.
In Fig.~\ref{fig:(14)}, the helicity amplitudes are presented at five angles, 
$\theta_{cm}=0^\circ, \ 45^\circ, \ 90^\circ, \ 135^\circ$, and $180^\circ$.
At $\theta_{cm}=0^\circ$, only $H_2$ has non-zero values, while
all the other elements vanish. In the backward direction, 
the non-vanishing element is $H_4$. Compared with the SAID analysis,
an overall agreement is obtained up to 500 MeV.

\subsection{{\it t}-channel vector meson exchange }

A long-standing question concerns the role
played by vector-meson exchange in low-energy pion 
photoproduction. According to the duality argument~\cite{dual}, 
the introduction of vector meson exchanges, 
along with a complete set of {\it s}-channel
resonances, might result in double counting. 
In practice, a systematic inclusion of all {\it s}-channel
resonances at the hadronic level is not available. 
Empirically, {\it t}-channel vector-meson exchange 
may account for incomplete {\it s}-channel resonance contributions,
which however makes the duality hypothesis more ambiguous.

Given the results presented in the previous Subsections,
the quark model framework, with an effective Lagrangian, 
could address this question in pion photoproduction
from a more fundamental level. 
As seen in the cross sections for $\pi^+$ production
up to 700 MeV, forward 
peaking above the Delta resonance has been successfully sustained 
up to 500 MeV. This could reasonably illustrate
that the effective Lagrangian has been sufficient to describe
the data over the first resonance region.
In order to consider the effect of possible double counting 
between the {\it t}-channel vector meson exchange and {\it s}-
and {\it u}-channel resonances, we compare models including various
subsets of these diagrams.  Our purpose is to clarify 
whether the behavior of those higher excited states (terms) would 
be similar to the inclusion of vector meson exchange,
particularly to compare with an `isobaric' model, 
where the {\it u}-channel resonance contributions are neglected. 

We shall introduce the following effective Lagrangians for vector-meson
exchange: 
\begin{equation}
{\cal L}_{\gamma\pi V}=e\frac{g_{\gamma\pi V}}{m_\pi}
\varepsilon_{\alpha\beta\gamma\delta}\partial^\alpha A^\beta\partial^\gamma
V^\delta \pi,
\end{equation}
for $\gamma\pi V$ coupling and 
\begin{equation}
{\cal L}_{Vqq}=g_{Vqq}\overline{\psi}\left(\gamma_\mu
+\frac{\kappa_q}{2m_q}\sigma_{\mu\nu}\partial^\nu\right)V^\mu\psi,
\end{equation}
for the quark-vector-meson ($V$-$qq$) coupling; 
$A$ and $V$ denote the photon and vector meson; $\pi$ denotes
the pion field 
and $\psi$ ($\overline{\psi}$) denotes the quark (antiquark) field; 
$g_{\gamma\pi V}$ and  $g_{Vqq}$ are coupling constants. 
Note that we treat the $V$-$qq$ coupling at quark level in order
to be consistent
with our framework. In this way, there is once again 
no need to introduce free parameters
for the vertex form factors. In addition, a simple current analogy
will relate the $g_{Vqq}$ to $g_{VNN}$, constraining this term 
as well.

Some simple algebra gives the transition amplitude,
\begin{eqnarray}
\label{vme}
{\cal M}_V&=&e\frac{g_{V\pi\gamma}g_{Vqq} 
e^{-({\bf k}-{\bf q})^2/6\alpha^2}}
{m_\pi (t-m^2_v)}
\left\{ g_t\left[1+\frac{\omega_m}{E_f+M_f}
+\frac{\omega_\gamma}{E_i+M_i}\right.\right.\nonumber\\
&&\left. +\frac{\kappa_q}{2m_q}\left(\frac{m^2_\pi}{E_f+M_f}
-\left(\frac{1}{E_f+M_f}+\frac{1}{E_i+M_i}k\cdot q\right)\right)\right]
{\bf q}\cdot({\bf k}\times\veps)\nonumber\\
&&+g_A
\left[\frac{\omega_\gamma {\bf q}^2}{E_f+M_f} +
\frac{\omega_m {\bf k}^2}{E_i+M_i} 
-\left(\frac{\omega_\gamma}{E_i+M_i}+\frac{\omega_m}{E_f+M_f}\right)
{\bf q}\cdot{\bf k} \right.\nonumber\\
&&+\frac{\kappa_q}{2m_q}\left(\omega_m {\bf k}^2 +\omega_\gamma {\bf q}^2
+\frac{\omega_\gamma\omega_m}{E_f+M_f}{\bf q}^2
+\frac{(\omega_\gamma\omega_m-m^2_\pi)}{E_i+M_i}{\bf k}^2\right.\nonumber\\
&&\left.\left. -\left(\omega_\gamma+\omega_m 
+\frac{\omega^2_m}{E_f+M_f}+\frac{\omega^2_\gamma}{E_i+M_i}
-\frac{k\cdot q}{E_i+M_i}+\frac{{\bf q}\cdot{\bf k}}{E_f+M_f}\right)
{\bf q}\cdot {\bf k}\right)\right]i\vsig\cdot\veps\nonumber\\
&&+g_A\left[\frac{\omega_m}{E_f+M_f}
+\frac{\kappa_q}{2m_q}\left(\omega_m+\frac{\omega^2_m}{E_f+M_f}
-\frac{k\cdot q}{E_f+M_f}+\frac{{\bf k}\cdot{\bf q}}{E_i+M_i}\right)\right]
i\vsig\cdot{\bf k}{\bf q}\cdot\veps \nonumber\\
&&\left. -g_A\left[\frac{\omega_\gamma}{E_f+M_f} 
+\frac{\kappa_q}{2m_q}\left(\omega_\gamma+\frac{{\bf k}^2}{E_i+M_i}
+\frac{\omega_\gamma\omega_m}{E_f+M_f}\right)\right]i\vsig\cdot{\bf q}
{\bf q}\cdot\veps
\right\} \ ,
\end{eqnarray}
where $k\cdot q=\omega_\gamma\omega_m-{\bf k}\cdot {\bf q}$
is the four-momentum product;
The exponent comes from the nucleon wave functions, which 
plays the role of a form factor; $g_A$ is the axial vector
coupling and defined in the quark model as Eq.~(\ref{ga}), i.e.
\begin{equation}
\langle N_f | \sum_j{\hat I}^v_j \vsig_j|N_i \rangle 
\equiv g_A\langle N_f | \vsig|N_i \rangle ,
\end{equation}
where ${\hat I}^v_j$ is the isospin operator for the exchanged vector meson.
The other factor $g_t$ comes from the isospin space,
\begin{equation}
g_t\equiv \langle N_f | \sum_j{\hat I}^v_j |N_i \rangle.
\end{equation}

Analogy between the quark level operator and $V$-$NN$ coupling
gives:
\begin{eqnarray}
g_t g_{Vqq}&=&g_{VNN}\nonumber\\
g_A\frac{g_{Vqq}}{m_q}(1+\kappa_q)&=&\frac{g_{VNN}}{m_N}(1+\kappa_V) ,
\end{eqnarray}
where $m_q=330$ MeV is the constituent quark mass. 
In Ref.~\cite{riska-01}, a similar relation was investigated, but only 
the vector current was introduced for the $V$-$qq$ coupling. 
We shall use the commonly used values for $g_{VNN}$ and $\kappa_V$
to constrain the values for $g_{Vqq}$ and $\kappa_q$.

In the $\pi^+$ production, we adopt the values $g_{\rho NN}=3$ 
and $\kappa_\rho=3.71$ as inputs. With the quark model values 
$g_A=5/3$ and $g_t=1$ for $\gamma p\to \pi^+ n$, 
we derive $g_{\rho qq}=3$ and $\kappa^\rho_q=-0.0064$.
In $\gamma p\to \pi^0 p$, we adopt $g_{\omega NN}=9$, 
and $\kappa_\omega=-0.12$. With $g^\omega_A=1$ and $g^\omega_t=3$,
$g_{\omega qq}=3$ and $\kappa^\omega_q=0.2$ are derived for the $\omega$
exchange, and with $g^\rho_A=5/3\sqrt{2}$ and $g^\rho_t=1/\sqrt{2}$,
$g_{\rho qq}=3$ and $\kappa^\rho_q=-1.99$ are derived.

\begin{figure}
\begin{center}
\epsfig{file=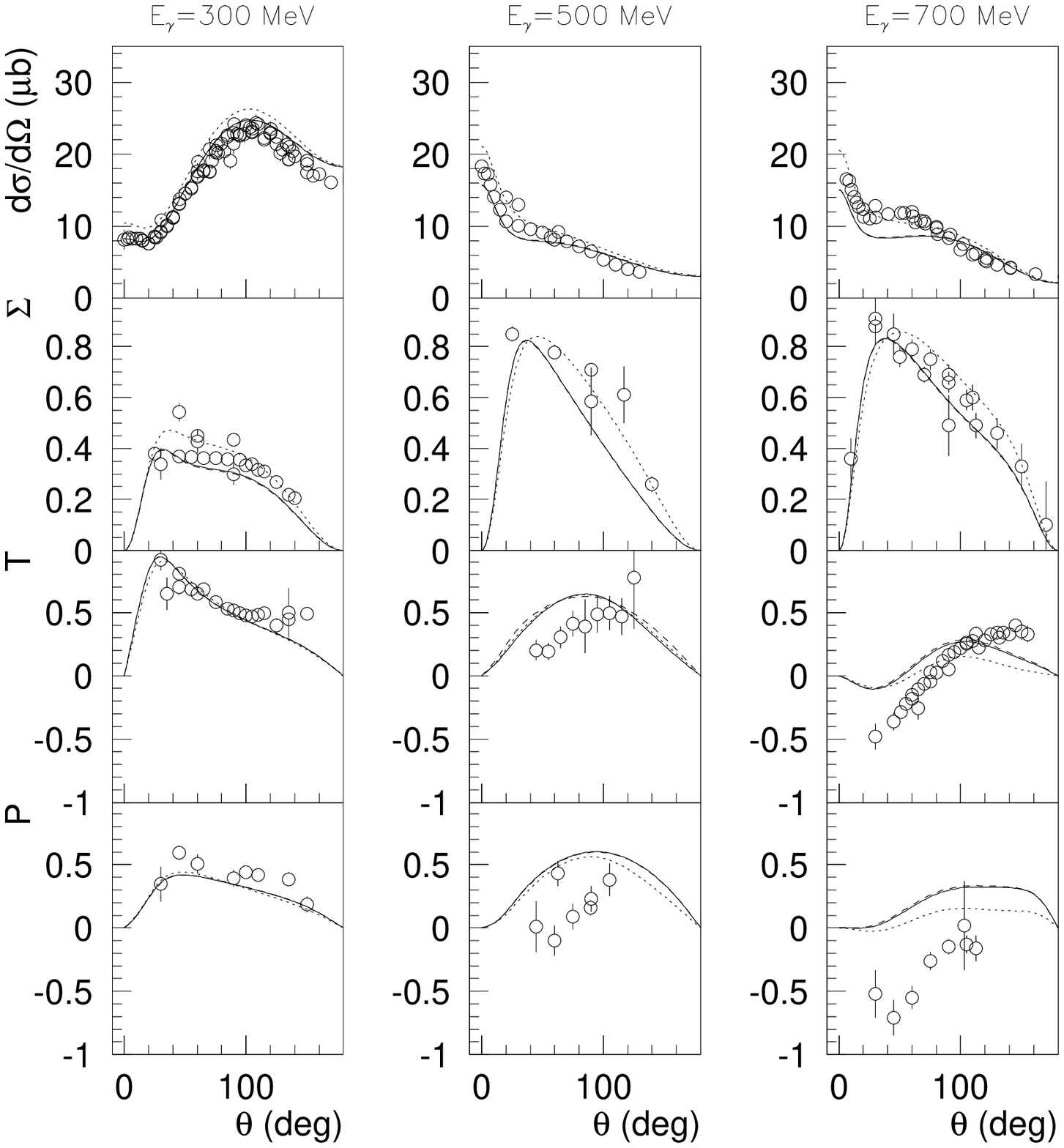,width=12.cm,height=12.cm}
\end{center}
\caption{ Observables for $\gamma p\to \pi^+ n$ at three energies.
Solid curves denote results for {\it s}+{\it u}; 
dashed for {\it s}+{\it u}+{\it t};
and dotted for {\it s}+{\it t}.
Data are from Ref.~\protect\cite{fu(71),da(01),bu(94),ec(67),ke(74),bs(79),as(72),du(96),al(76)1,eg(81)}.}
\protect\label{fig:(15)}
\end{figure}

\begin{figure}
\begin{center}
\epsfig{file=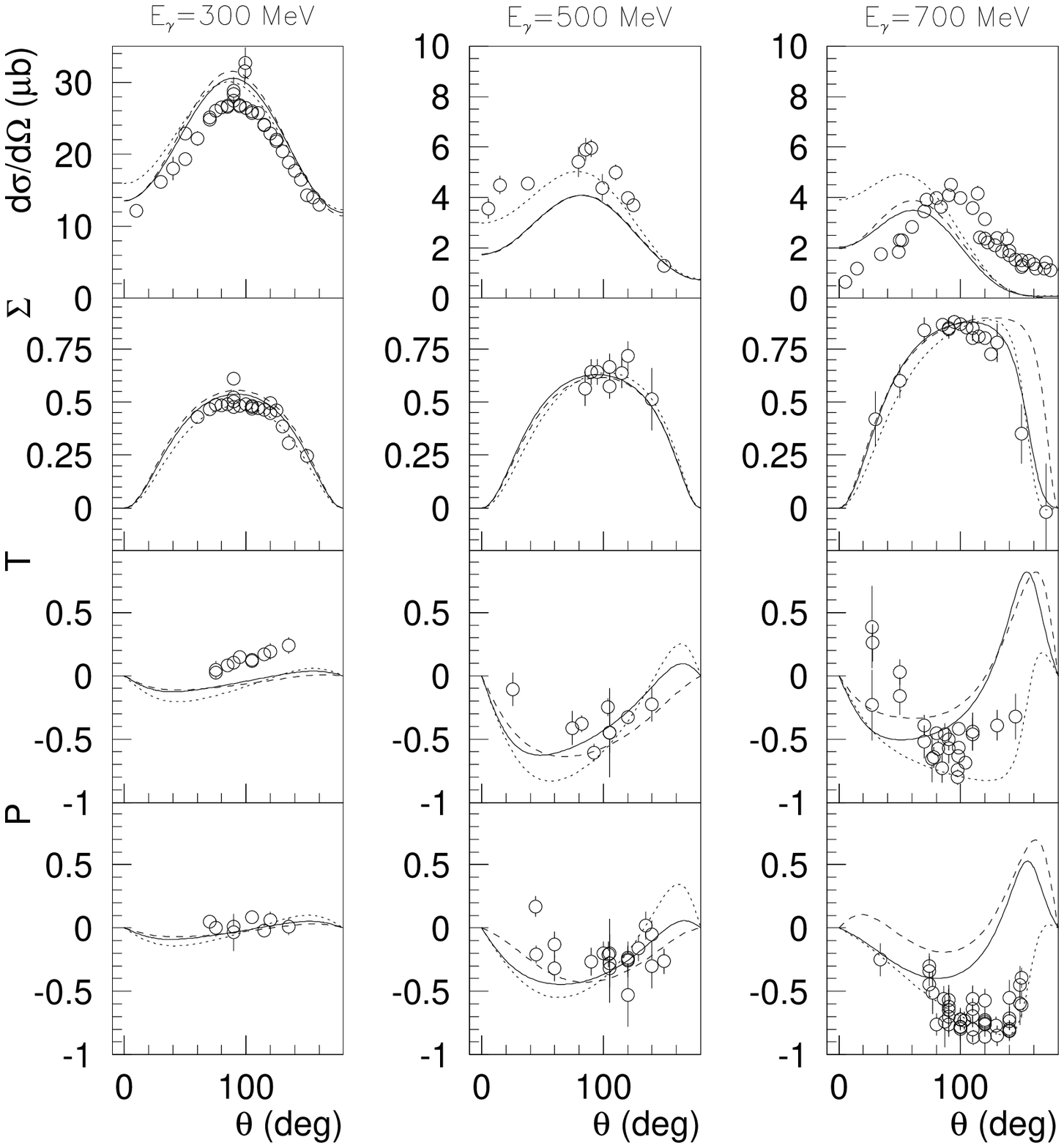,width=12.cm,height=12.cm}
\end{center}
\caption{ Observables for $\gamma p\to \pi^0 p$ at three energies.
Solid curves denote results for {\it s}+{\it u}; 
dashed for {\it s}+{\it u}+{\it t};
and dotted for {\it s}+{\it t}.
Data are from Ref.~\protect\cite{kr(99),yo(80),do(75),al(79),hy(73),ke(74),ad(01),bh(77),fe(76),fk(78),zy(78),ka(80)}.}
\protect\label{fig:(16)}
\end{figure}

In Figs.~\ref{fig:(15)} and ~\ref{fig:(16)}, 
we show the calculations of observables 
($d\sigma/d\Omega$, $\Sigma$, $T$ and $P$) 
with the {\it t}-channel vector meson exchange (VME) for $\gamma p\to \pi^+ n$
and $\gamma p\to \pi^0 p$, respectively.
Three energy bins are investigated.
We use {\it s}+{\it u} to denote the effective Lagrangian
calculations, while {\it s}+{\it t} denotes calculations 
suppressing the {\it u}-channel resonance but including the {\it t}-channel
VME. Finally, we use {\it s}+{\it u}+{\it t} to represent the full calculation,
including the VME. The values, 
$g_{\rho\pi\gamma}=0.103$ and $g_{\omega\pi\gamma}=0.313$ are adopted.

In the $\pi^+$ production, contributions from the VME are found to be
negligible. One reason is the relatively smaller 
couplings of $g_{\rho\pi\gamma}$ and $g_{\rho NN}$ compared 
to the couplings for $\omega$ exchange. 
However, the main factor leading to small VME contributions
in the $\pi^+$ production is a large cancellation
occurring among terms proportional to $g_A$ and $g_t$ in Eq.~(\ref{vme}).
Since different contributions from the VME to different
reactions depend on the quark model prediction for $g_A$ and $g_t$,
the VME might introduce more model-dependent ingredients
in the calculations.

\section{Conclusions}

We have studied pion photoproduction in four charge channels
within the quark model incorporating an effective Lagrangian. 
Up to $E_\gamma=$ 500 MeV, the cross sections and 
single polarization asymmetries can be accounted for 
with one adjustable parameter for the Delta excitation strength. 
We find that if a stronger coupling
for the Delta transition is employed, all the observables 
in the first resonance region can be reproduced. 
In another words, the NRCQM might have provided a reasonable 
form factor for the Delta resonance, but with weaker
coupling. 

As the first systematic microscopic study of pion photoproduction, 
this result suggests that a direct calculation of the tree level 
diagrams based on the quark model with a chiral effective Lagrangian
may contain the main ingredients required in 
an elementary approach. In particular, we show that 
the quark model form factors play a key role in reproducing 
the data over a wide energy region. 
Such a form factor can be only self-consistently and completely
considered in a direct calculation of quark level diagrams.
This result highlights the relation between the background
Born terms and resonance excitations. Extensions to higher
energies would help to clarify the relation between quark model
results and those found via phenomenology. 
Nevertheless, a parallel investigation of electroproduction
would be useful for a better understanding of the Delta 
resonance based on this model. We shall report this elsewhere.

Restricted to the low energies at $E_\gamma < 500$ MeV,
we see that {\it t}-channel vector-meson exchange
is negligible. This leaves the duality hypothesis far from 
conclusive. We expect that more sophisticated calculations
at high energies could be helpful in disentangling this question
as well.

\acknowledgments

Useful discussions with F.E. Close and T. Sato are gratefully 
acknowledged. Q.Z. thanks J. Arends for sending details about 
their experimental data.
Q.Z. and J.A. acknowledge 
the financial support of the U.K. Engineering and Physical 
Sciences Research Council (Grant No. GR/M82141).
R.W. acknowledges the financial support in part by a U.S. 
Department of Energy Grant DE-FG02-99ER41110. R.W. also
acknowledges partial support from Jefferson Lab, by the
Southeastern Universities Research Association under
D.O.E. contract DE-AC05- 84ER40150.


%
%
%
%
%
%
%
%
%
%
%

%

%

%


\begin{references}
\bibitem{born-delta} Plotted is a SAID analysis containing the unitarized
Born terms plus Delta multipoles found in a fit to data (up to 2 GeV).
\end{references}
\end{document}